\documentclass[twocolumn]{aastex63}

\usepackage{textcomp}
\usepackage{graphicx}
\usepackage{amssymb}
\usepackage{amsmath}
\usepackage{epstopdf}
\usepackage{gensymb}
\usepackage{natbib}
\usepackage{appendix}
\usepackage{tabu}
\usepackage{color}
\usepackage{multirow}

\usepackage{booktabs}
\usepackage{rotating}

\newcommand{\ca}{\mbox{Ca\,{\sc ii}~K\,~}}
\newcommand{\cahk}{\mbox{Ca\,{\sc ii}~H \&~K\,~}}
\newcommand{\cah}{\mbox{Ca\,{\sc ii}~H\,~}}
\newcommand{\fe}{\mbox{Fe\,{\sc i}\,~}}

\begin{document}
\title{Investigating the effect of solar ambient and data characteristics \\
on \ca observations and line profile measurements}
\received{\today}

\submitjournal{ApJ}
\shortauthors{Murabito et al.}

\correspondingauthor{M.~Murabito}
\email{mariarita.murabito@inaf.it}

\author[0000-0002-0144-2252]{M.~Murabito}
\affiliation{INAF Osservatorio Astronomico di Capodimonte, Salita
Moiariello 16, 80131 Naples, Italy}

\author[0000-0003-2596-9523]{I.~Ermolli}
\affiliation{INAF Osservatorio Astronomico di Roma, Via Frascati 33, 00078 Monte Porzio Catone, Italy}

\author[0000-0002-0335-9831]{T.~Chatzistergos}
\affiliation{Max Planck Institute for Solar System Research, Justus-von-Liebig-Weg 3, 37077 G\"{o}ttingen, Germany}

\author[0000-0002-7711-5397]{S.~Jafarzadeh}
\affiliation{Max Planck Institute for Solar System Research, Justus-von-Liebig-Weg 3, 37077 G\"{o}ttingen, Germany}
\affiliation{Rosseland Centre for Solar Physics, University of Oslo, P.O. Box 1029 Blindern, 0315 Oslo, Norway}

\author[0000-0002-0974-2401]{F.~Giorgi}
\affiliation{INAF Osservatorio Astronomico di Roma, Via Frascati 33, 00078 Monte Porzio Catone, Italy}

\author[0000-0003-2088-028X]{L.~Rouppe van der Voort}
\affiliation{Institute of Theoretical Astrophysics, University of Oslo, PO Box 1029, Blindern 0315, Oslo, Norway}
\affiliation{Rosseland Centre for Solar Physics, University of Oslo, P.O. Box 1029 Blindern, 0315 Oslo, Norway}

\begin{abstract}
We analysed state-of-the-art observations of the solar atmosphere to investigate the dependence of the \ca brightness of several solar features on spectral bandwidth and spatial resolution of the data. Specifically, we study data obtained at the Swedish Solar Telescope with the CRiSP and CHROMIS instruments. The analyzed data, which are characterized by spectral bandwidth of 0.12 \AA \ and spatial resolution of 0.078\arcsec, were acquired close to disc center by targeting a quiet Sun area and an active region. We convolved the original observations with Gaussian kernels to degrade their spectral bandwidth and spatial resolution to the instrumental characteristics of the most prominent series of \ca observations available to date. We then studied the effect of data degradation on the observed regions and on parameters derived from \ca line measurements that are largely employed as diagnostics of the solar and stellar chromospheres. We find that the effect of degrading the spectral resolution of \ca observations and line profiles depends on both the employed bandwidth and observed solar region. Besides, we found that the spatial degradation impacts the data characterized by a broad bandwidth to a larger extent compared to those acquired with a narrow band. However, the appearance of the observed solar regions is only slightly affected by the spatial resolution of data with bandwidths up to 1 \AA \ and in the range [3,10] \AA. Finally, we derived relationships that can be used to intercalibrate results from observations taken with different instruments in diverse regions of the solar atmosphere. 

\end{abstract}


\section{Introduction}
\label{sec1}
Solar observations have often served as benchmarks of stellar conditions \citep{Schmelz2003,Engvold2019}.  
A particularly illustrative example of the above link is given by
the observations in the \cahk lines at 
3968.47~\AA~and 3933.67~\AA, respectively, which are 
the two deepest and broadest absorption lines in the visible spectrum of the Sun. 

In fact, early observations of the solar disc at the cores of the \cahk lines revealed  brightenings in large regions surrounding sunspots and in a network pattern across the whole solar disc \citep{Hale1903}. Concurrently to these observations,  
late-type stars were also found to commonly show emissions at the \cahk lines and thus considered to have atmospheric layers similar to those of the Sun  
\citep{Eberhard1913}. Later on, solar observations revealed a clear association between brightening at the \cahk lines and magnetic field strength \citep{Babcock1955,Howard1959,Chatzistergos2019b} and area \citep{Leighton1959,Sheeley1967}. Along with previous observations, this association allowed  the \cahk emissions to be used as an indicator of the  magnetic fields in the Sun and other stars \citep{Wilson1978}.  
Since then, measurements at the \cahk lines have widely been used to trace changes in the surface 
of the Sun and other stars due to magnetic activity and other processes such as rotation and convection  
\citep[e.g.,][and references therein]{White1978,Keil1984,Noyes1984, Baliunas1984,Baliunas1995, HallLockwood1995,Radicketal1998,Hall2007,Radick2018}.  

Furthermore, for many years, the \cahk lines have been used as one of the most reliable diagnostics of the physical properties of the solar \citep{Linsky1970,Linsky1970b} and stellar 
\citep{Linsky2017} chromospheres. This is due to the characteristic profile 
of the \cahk lines, which show two peaks and two secondary minima 
towards the violet or red part of the spectrum relative to the line centre,  
and a reversal at the line centre. These line features result from emissions originating from the photosphere, where the temperature decreases with height until a temperature minimum is reached, to the overlying lower chromosphere, where the temperature increases with height \citep{linsky1968}. 
Following the notation given by \citet{Hale1903}, in the \ca line the abovementioned line features are labeled as K$_\mathrm{1V}$, K$_\mathrm{1R}$, K$_\mathrm{2V}$, K$_\mathrm{2R}$, and K$_3$, respectively. Noteworthy, they all occur within a 1~\AA \ interval and are qualitatively the same for both the quiet Sun and plages regions. However, for the latter the K$_\mathrm{2V}$ and K$_\mathrm{2R}$ peaks and the K$_{3}$ minimum display a significant intensity increase 
relative to the quiet Sun \citep{Linsky1970}. 

It is worth noting that, although both spectral regions are observable from the ground, the \cahk lines have not been equally explored, with existing observations in favour of the \ca line. Contributing to this is that the \cah line is blended with the Balmer H$\epsilon$ line at 
3970.07~\AA. Another reason is that \cah line is slightly less sensitive than \ca as an atmospheric diagnostic. That is because  the \cah emission peaks are often less pronounced than in the \ca ones, thus probing lower heights in the chromosphere. Further information on the formation and diagnostic potential for the solar atmosphere of the \cahk lines can be found in, e.g.,  \citet{Linsky1970} and \citet{bjorgen2018}.    

The literature concerning solar and stellar observations at the \ca line is extensive and  suggestive of plenty of data available at that radiation. 
In fact, full-disc spatially-resolved solar observations at the \ca line have continuously been performed since 1892 with various telescopes operating at the \ca line with bandwidths in the range [0.09,10]~\AA \ centered at the line core, and spatial resolution larger than 1\arcsec \ 
\citep[e.g.,][]{Chatzistergos2018,Chatzistergos2019a,Chatzistergos2020,chatzistergos2022}. 
Most prominent archives of historical \ca data are e.g. those of the Meudon \citep{malherbe_potential_2022}, Kodaikanal \citep{chatzistergos2019c}, Mt Wilson \citep{lefebvre_solar_2005}, and Coimbra \citep{lourenco_solar_2019} Observatories, while the ones for modern data are e.g. those of the Rome \citep[][]{ermolli2022fass} and Kanzelh\"ohe \citep[][]{poetzi2021} Observatories. 
Besides, since late 1960s the \ca line emission integrated over the solar disc has been measured almost daily at e.g. the Kitt Peak and Sacramento Peak sites of the US National Solar Observatory \citep[e.g.,][]{whitelivingston1998,scarglekeil2013}, and at the Kodaikanal Observatory in India \citep[e.g.,][]{sivaraman1987}.   
Furthermore, there are also \ca observations at high spectral resolution, on the order of  0.2~\AA \ or better, acquired on spatially resolved regions of the solar disc with a spatial resolution even better than 0.2\arcsec, but over limited time-intervals and disc positions. Such data are e.g. those obtained over the last few years at the Swedish Solar Telescope, which are further described in the following. 

In addition to the observations described above, since 1960s the disc-integrated \ca emission of the Sun and late-type stars has been monitored at the Mt Wilson Observatory \citep[1966--2003,][]{Wilson1978,Duncan1991,Baliunas1995}, then at the Lowell Observatory  
\citep[1994--present,][]{Hall2007}, and more recently with the Potsdam Echelle Polarimetric and Spectroscopic Instrument (PEPSI) of the Large Binocular Telescope \citep[LBT, see e.g.][]{dineva2022}, with the HARPS-N spectrograph at the Telescopio Nazionale Galileo \citep[TNG, see e.g.][]{maldonado2019}, and with the photometers on-board e.g. the CoRoT \citep{Michel2008, Auvergne2009,Gondoin2012} and KEPLER \citep{kepler2010,kepler20102} missions. 

It is worth noting that, in spite of being widely observed over many years, several aspects of the \cahk emissions in atmosphere of the Sun and other stars are still not fully understood. Examples are the link between the coverage of a stellar disc by magnetic features and the \cahk emissions \citep[see, e.g.,][]{Sowmya2021}, and the relationship between \cahk brightening and magnetic field strength \citep[see, e.g.,][]{Chatzistergos2019b}. These gaps in the knowledge of the \cahk emissions depend on several factors. Firstly, all the available observations report relative photometric data with respect to some standard that is subjective of a biased definition. For example, the disc-resolved solar observations carry information of the brightening at the \ca line with respect to an average quiet-Sun emission, whose definition depends on the data characteristics and their processing methods. Likewise, the disc-integrated  measurements describe line parameters with respect to a continuum reference that is hardly identified in the solar and stellar spectra adjacent to the \cahk lines, because of the presence of many absorption lines. Unfortunately, the various observations that are available in the literature have often been obtained with diverse instruments and methodologies, which renders their comparative analysis inconclusive.  

We aim to contribute to a better knowledge of the \cahk emissions, in particular of the relationship between \ca emission and magnetic field strength. As a first step, here we investigate the dependence of the \ca line data on different ambient conditions of the solar atmosphere due to different levels of magnetic flux, and characteristics of the observations, by using state-of-the-art data taken at solar disc center. 

The paper is structured as follows. In Section \ref{sec2} we describe the data analysed in our study and the methods used to process them. In Section \ref{sec3} we present our results on the measured \ca emissions depending on the spectral and spatial resolution of the analysed observations. We discuss our results and summarise them by drawing our conclusions in Sections \ref{sec4} and \ref{sec5}, respectively.
	
\section{Data and Methods}
\label{sec2}
\subsection{Observations}

The data analysed in our study were acquired at the Swedish 1-m Solar Telescope  \citep[SST,][]{Scharmer2006} with the CRisp Imaging SPectropolarimeter  \citep[CRISP;][]{Scharmer2008} and with the CHROMospheric Imaging Spectrometer \citep[CHROMIS;][]{Scharmer2017} at the \fe doublet lines at 6301.51--6302.50~\AA \ (hereafter  
6301--6302~\AA) and the \ca line at 
3933.67~\AA \ (hereafter 3933~\AA), respectively. In particular, we analysed three series of simultaneous full-Stokes \fe and Stokes-I \ca observations acquired close to disc center by targeting a quiet-Sun area (QS) and an active region (AR), namely AR NOAA 12585. The CRISP and CHROMIS data are characterized by an image scale of $\approx$ 0.060\arcsec/pixel and 0.039\arcsec/pixel, and a spectral resolution of about 60~m\AA \ and 120~m\AA, respectively, over a Field-of-View (FoV) of about 1\arcmin$\times$1\arcmin. 

The QS data were acquired on 25 May 2017 from 08:06 to 11:16 UT at $\mu$=0.99, by targeting a region with patches of regular and irregular granulation, hereafter referred to as quiet-Sun granulation (QG) and quiet-Sun magnetized granulation (QM), respectively.  
The QS data consists of several scans of full-Stokes \fe and Stokes-I \ca measurements. The \fe line data were taken sequentially at 9 spectral positions around the \fe 6301~\AA \ line centre, at  6 spectral positions around the \fe 6302~\AA \ line centre, and 
at -0.26~\AA~from the \fe 6302~\AA~ line centre for the continuum, by referring to rest-frame line centres. 
The \ca data were taken at 41 equally spaced positions in the range [-1.30,+1.30]~\AA  \ relative to the line centre and at 4000~\AA \ for the reference continuum measurement. The cadence was 19.6~s and 13.6~s, for \fe and \ca respectively.

The AR data were acquired on 5 September 2016 from 09:48 to 10:07 UT
at $\mu$=0.99, by targeting a sunspot (SP) with umbral (UM) and penumbral (PE) regions and a neighbouring area (NA) with plages (PL) and  several pores (PO), respectively. Similarly to the QS set, the AR data include several series of full-Stokes \fe measurements and Stokes-I \ca data. Here the \fe doublet lines were sampled as 
described above for the QS observations, while the \ca line measurements were obtained at the following 21 spectral positions: [-1.4, -0.78, -0.70, -0.63, -0.55, -0.47, -0.39, -0.23, -0.16, -0.08,
0., +0.08, +0.16, +0.23, +0.39, +0.47, +0.55, +0.63, +1.25]~\AA~ relative to line center, i.e. on a slightly smaller spectral range than that used for  the QS data, and at 4000~\AA \ as the reference continuum. The \fe and \ca datasets were taken at cadence of 32 and 14 s, respectively.

The data analysed in our study were extracted from all the observations available for the QS, NA, and SP regions, as the best frames among the 20 non consecutive observations of each solar target that are   characterized by the highest contrast for best seeing during their acquisition.  

The QS, NA, and SP observations have been previously studied by \citet{Bose2019}, \citet{murabito2021}, and \citet{Pozuelo2019}, respectively. 

\subsection{Data reduction}

\begin{figure}
\centering
{
\includegraphics[scale=0.98, trim=10 10 0 0]{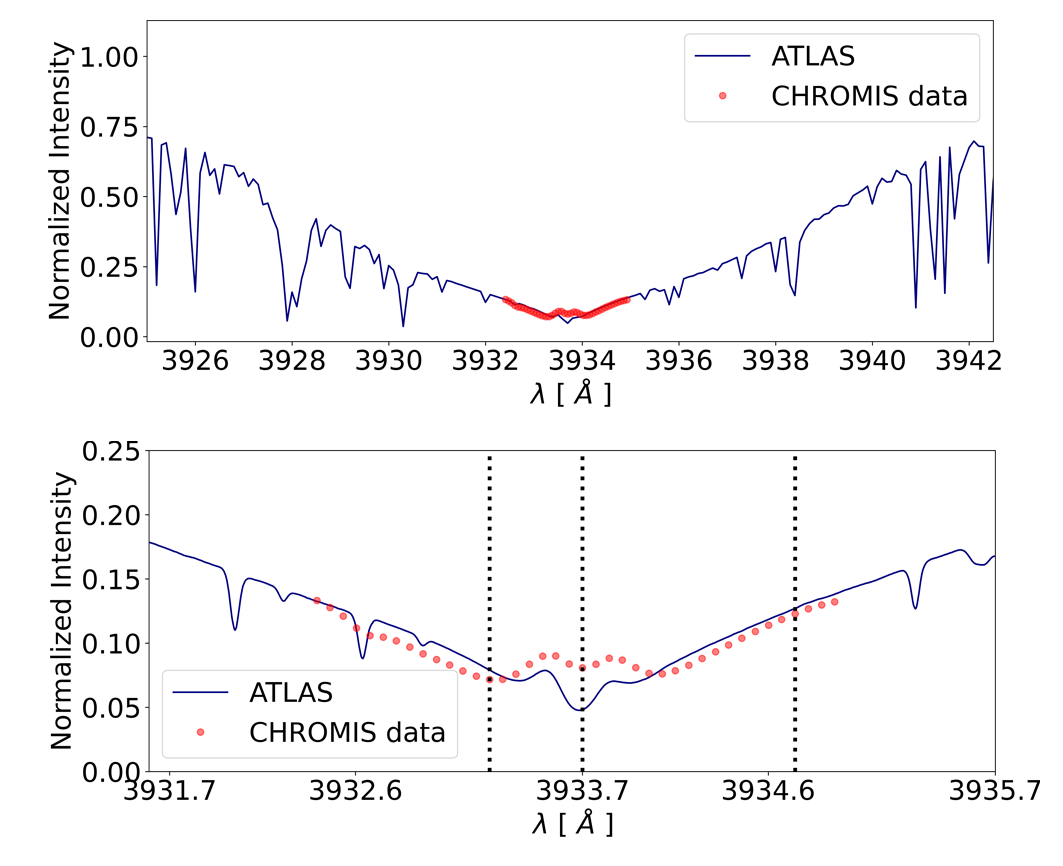}
}
	\caption{Examples of 
 CHROMIS 
 spectral sampling (red dots) of the 
 \ca  line 
 3933~\AA. 
 The measurements refer to a QS region and are normalized to the data at the reference continuum (not shown). 
 Overplotted are disc center atlas measurements (blue solid line) from \citet{Delbouille1973}. 
 Vertical dotted lines mark the spectral samplings of the \ca observations shown in Fig.~\ref{fig2}.
	}
	\label{fig1}
\end{figure}

The data  analysed in our study were reduced by using the CRISPRED \citep{delacruz2015} and CHROMISRED \citep{Lofdahl2021} pipelines for CRISP and CHROMIS data, respectively. The processing consists of several steps, including among others application of bias and flat-field response of the detectors, compensation for spatial variation in the spectral instrumental response, for optics aberrations, and for atmospheric turbulence with the Multi-Object Multi-Frame Blind Deconvolution \citep[MOMFBD,][]{vanNoort2005,Lofdahl2002} image restoration. The latter step compensates for residual seeing degradation over the FoV unaccounted for by the adaptive optic system of the SST. 

The data produced by the above pipelines were further processed as follows. 
Firstly, they were derotated to account for diurnal field rotation. Then we aligned the CRISP and CHROMIS data and reduced them to the same pixel scale, by using as reference the frames acquired at the \fe 6302~\AA~Stokes-I line continuum and \ca continuum at 4000~\AA, which both originate in the low photosphere. 
In particular, we scaled up the CRISP data to the pixel scale of the CHROMIS data. This was done to maintain the original resolution of the \ca observations. 
Then, we rotated the CRISP images and trimmed them to match the FoV of the corresponding CHROMIS images. 
We used a cross-correlation technique to apply vertical and horizontal image shifts that result in a close match at sub-pixel accuracy.  
The above processing led to images that cover a region of about 50\arcsec~$\times$~40\arcsec~ for each target, with images having slightly different dimension of 1369~$\times$~1066, 1367~$\times$~1004, and 1403~$\times$~1067 pixels$^2$ for the QS, NA, and SP datasets, respectively.

Figure \ref{fig1} shows examples of the spectral-line profiles extracted from the data obtained from the processing described above. In particular, we show the sampling of 
\ca 3933~\AA \ line  resulting from spatial averaging of the data over the whole FoV of the QS set (red dots), depicted on the full spectra (solid line) from the disc-center atlas measurements from \citet{Delbouille1973}\footnote{Available at the BASS2000 Archive.
https://bass2000.obspm.fr} for the sake of clarity. The data are normalized to the intensity of the reference continuum. We note that the spatially averaged \ca spectra are slightly higher  for QS than the reference. This could be due to the analysed FoV. 

Figure \ref{fig1} makes it clear that the available 
\ca data only probe the \ca line centre with its typical double reversal and the innermost part of the line wings, 
but they do not sample the extended \ca  line wings away from line centre that sample the deeper layers of the photosphere. 

To account for these characteristics of the data we applied the following processing  steps. 
First 
we performed the absolute wavelength and intensity calibration of the \ca observations. 
Due to the different shape of the profiles measured in the QS regions with respect to the ones measured in NA and SP areas\footnote{The different shape is mostly in terms of the line width close to the line centre.} we used, as reference, the solar disc center atlas data by \citet{Delbouille1973} for the QS observations, and the sunspot umbral spectra from \citet{KittPeakAtlas}\footnote{Available at https://nispdata.nso.edu/ftp/pub/atlas/spot4atl/ } for the NA and SP observations. 
Then, we extrapolated the analysed data to cover a wider spectral range than the one of the measured values, by using the atlas measurements at several spectral positions in the line wings as a reference. In particular, we assumed that the values extrapolated for the measured series follow the ones in the atlas at the following spectral positions [3916.9, 3924.0, 3929.6, 3932.4, 3935.0, 3939.5, 3942.1, 3949.3]~\AA \ for the QS observations, and at the following spectral positions [3924.0, 3929.6, 3932.4, 3935.0, 3939.5, 3942.1, 3949.3]~\AA \ for the NA and SP observations. It is worth noting that these positions were selected after several tests and trials aimed at optimizing the results obtained, while the slightly different 
spectral positions considered for the QS and other sets derive from the fact that the measurements of the atlas umbral spectra are only available for wavelengths larger than 3920.5~\AA. 

We used results from linear interpolations between pairs of the above reference positions to reconstruct values of an extrapolated line profile that mimic the behaviour of emission in the \ca line wings.  
However, we note that the data obtained by combining the measured and extrapolated values simulate high resolution spectral observations of the \ca line centre and low resolution spectral observations of line wings, the latter without accounting for the many lines that populate the spectra adjacent to the \ca  line centre. We further consider this limitation of the data obtained from our processing in the following.  

Finally, we applied spectral and spatial degradation to the data obtained from the previous processing steps, by convolving them with Gaussian kernel functions of varying width, in order to investigate the effect of spectral bandwith and spatial resolution of the data on \ca observations and line measurements of various solar features. In particular, we convolved the CHROMIS observations, which were taken with a 0.12~\AA \ spectral bandwidth, with 1D Gaussian functions having full-width-half-maximum (FWHM) of [0.2, 0.3, 0.4, 0.5, 0.6, 0.7, 1.0, 1.8, 2.5, 3.0, 5.0, 10.0]~\AA. These spectral widths (hereafter referred to as bandwidths and spectral degradations) match the bandwidths of most of the existing series of full-disc solar observations at the \ca \citep[see e.g. Tables 1 and 2 in ][]{chatzistergos2022fass}. Moreover, we convolved the original data (with a spatial sampling of 0.039\arcsec/pixel) with a spatial 2D Gaussian kernel with FWHM of [0.18, 0.3, 1.0]\arcsec. This spatial degradation was to represent the spatial resolution of the  photospheric and chromospheric   observations acquired with the {\sc Sunrise}'s Imaging Magnetograph eXperiment \citep[{\sc Sunrise}/IMaX,][]{2010ApJ...723L.127S,barthol2011,pillet2011,2017ApJS..229....2S}, Hinode's Solar Optical Telescope \citep[Hinode/SOT,][]{tsuneta2008,ichimoto2008}, and SDO's Helioseismic Magnetic Imager \citep[SDO/HMI,][]{pesnell2012,scherrer2012,schou2012}, respectively. We note that at present, the data from the above instruments are the most widely employed for studies of the photosphere and chromosphere. In addition, we also investigated the full-disc solar observations produced with a moderate spatial resolution resulting from a  pixel scale larger than 2\arcsec/pixel. As mentioned above, these observations have been regularly obtained  at several observatories since the beginning of the 20th century. We note that the instance of these observations is  interesting in light of the role they play in connecting series of historical and modern full-disk solar \ca line observations  \citep{chatzistergos2022fass}. In this case we convolved the original CHROMIS data with a spatial 2D Gaussian kernel having FWHM of 4\arcsec.  

On both the original observations and data obtained from the above processing (i.e., the degraded data), we then evaluated two observable parameters of the \ca line that are widely employed for the monitoring of the chromosphere on the Sun and late-type stars. In particular, following, e.g., \citet[][]{scarglekeil2013,bjorgen2018,dineva2022}, we estimated the K$_3$ intensity in \ca line core and the emission index equivalent width in 
1~\AA \ band centered on the line profile (hereafter referred to as $EMDX$). Indeed, among the various \ca line parameters employed in the literature, K$_3$ and $EMDX$ are the ones most sensitive to changes of the \ca line profile. We computed K$_3$ by measuring the  intensity at 3933.67 \AA, and $EMDX$ by integrating the data with the five-point Newton-Cotes integration formula of the \textit{int$\_$tabulated} function in the \textit{Interactive Data Language (IDL)}.

\section{Results}
\label{sec3}

\begin{figure*}
\centering{
    \includegraphics[scale=0.89,trim=0 10 0 0,clip]{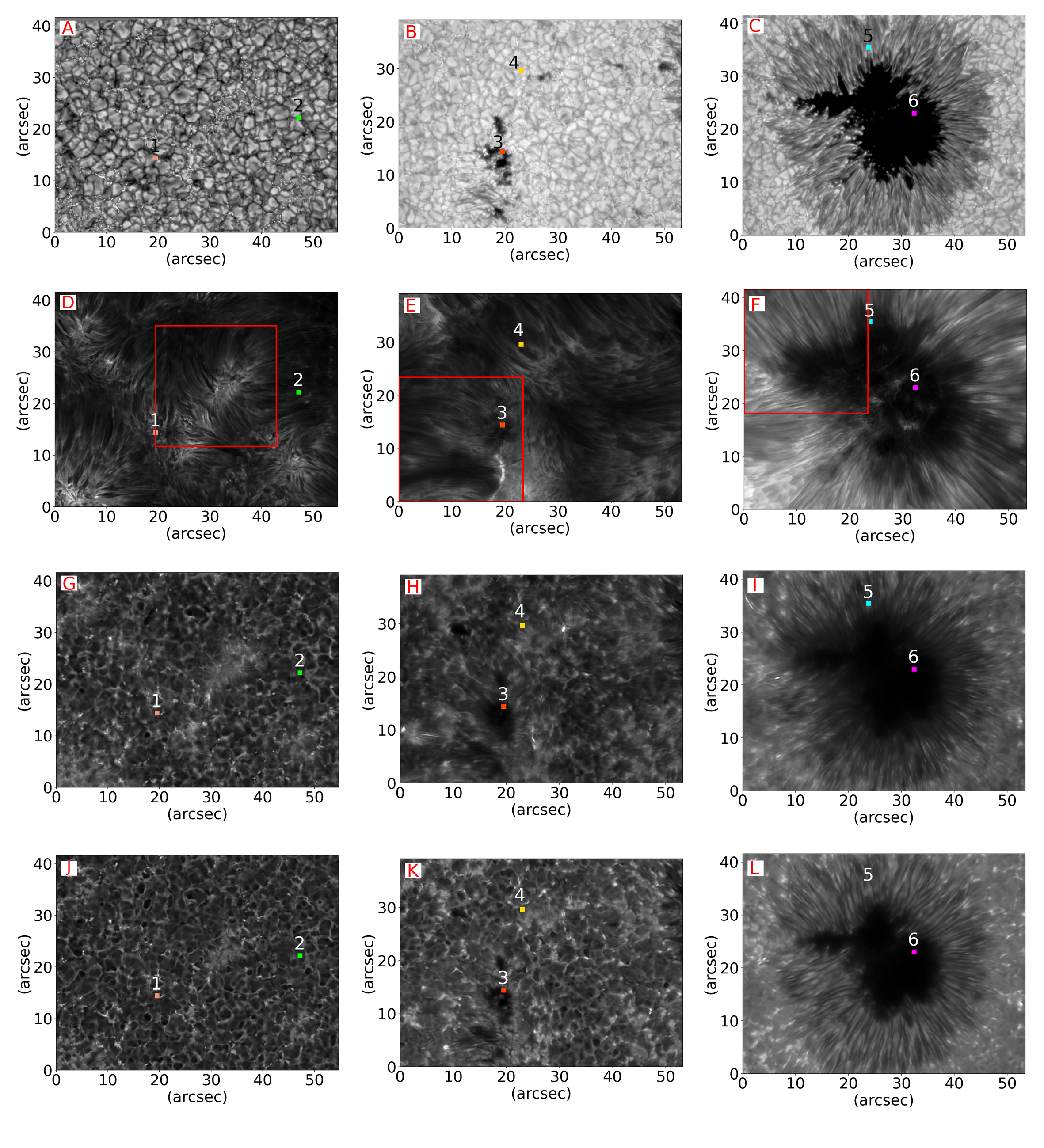}
}
	    \caption{CRISP \fe 6302~\AA~continuum images (panels 
     A--C) and CHROMIS \ca 3933~\AA \  images in the line core (panels D--F), at  
	    -0.42~\AA \ from the line core at roughly K$_\mathrm{1V}$ (panels G--I), and at 
	    +1.05~\AA \ in the red wing (panels J--L) of the three regions analysed in our study, referring to a quiet Sun area (left column panels), an active  region with plages and several pores (middle column panels), and a sunspot with umbra and penumbra (right column panels). 
	    After the image processing described in Sect. 2.2, the CRISP and CHROMIS observations are shown here with same dimension and pixel scale, which in the original data are in favour of the CHROMIS observations and is maintained here. Each  observation is  shown using the  intensity interval that enhances the visibility of the solar features therein. 
	         The small coloured boxes show  the six 
          20~$\times$~20 pixels wide areas randomly selected in the observations to represent the solar features analysed in our study in the ambient to which they belong. The boxes are shown in all panels to make comparisons easier.  The red boxes on line core observations mark the regions shown in Figs. \ref{fig5}, \ref{fig7}, and in Appendix B.
      	    }
	\label{fig2}
\end{figure*}

\begin{figure*}
\centering{
\includegraphics[scale=0.9,trim=0 0 0 50]{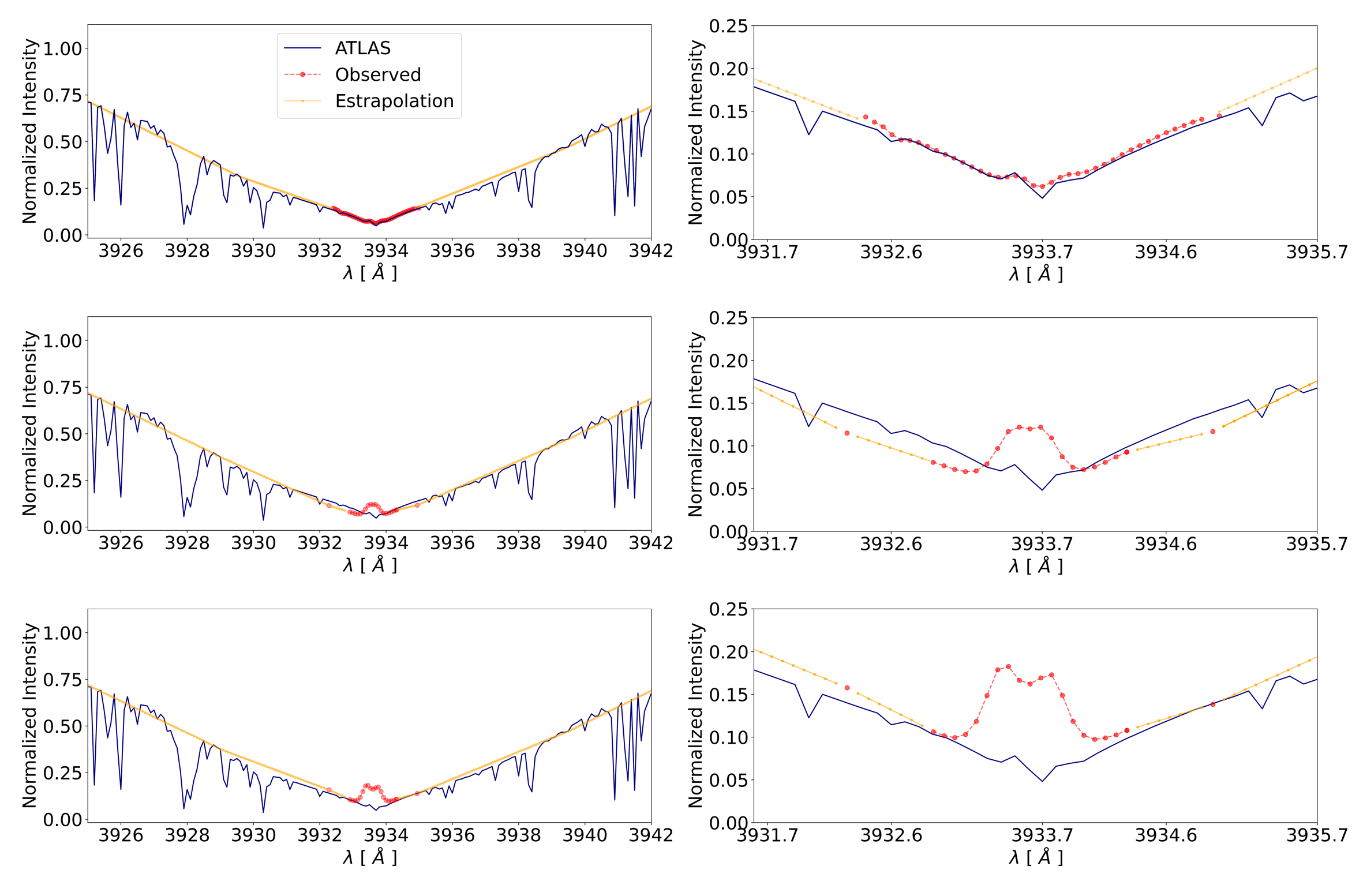}
}
	\caption{Examples of \ca line profiles analysed in our study. These were obtained by averaging data in sub-arrays representative of quiet areas in the studied QS (top panels), NA (middle panels), and SP (bottom panels) regions. The values are normalized to the data at the reference continuum at 4000~\AA \ (not shown). The blue profile and red dots in each panel indicate the atlas data from \citet{Delbouille1973} and our observations, respectively. The yellow line describes the results from the data extrapolation based on atlas values.
	}
	\label{fig3}
\end{figure*}

\begin{figure}
\centering{
\includegraphics[scale=0.98, trim=10 10 0 0]{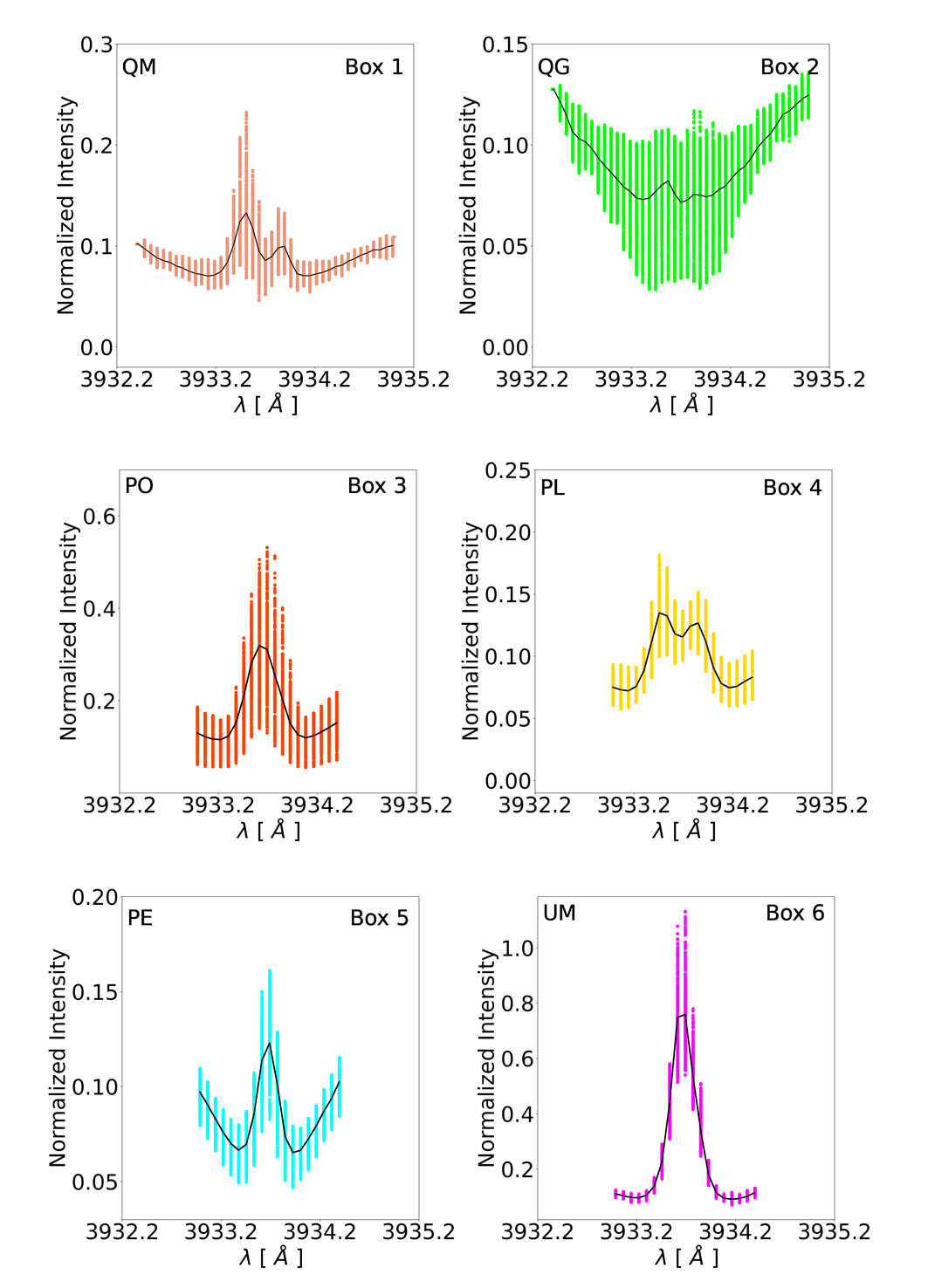}
  }
  \caption{Examples of \ca line profiles analysed in our study. These were obtained from the 20~$\times$~20 pixels wide areas marked with numbered and coloured boxes in Fig.~\ref{fig2}. Black solid lines show the mean profiles derived from the spatially-resolved data available at each observed spectral position. We recall that the QS region (top row panels), to which the QM and QG areas belong, was sampled over a slightly larger spectral range than the NA (middle row panels) and SP (bottom row panels) regions, where the PO, PL, PE and UM areas were selected.
	}
	\label{fig4}
\end{figure}

\begin{figure*}
\centering{
	\includegraphics[scale=0.87, trim=10 10 0 0]{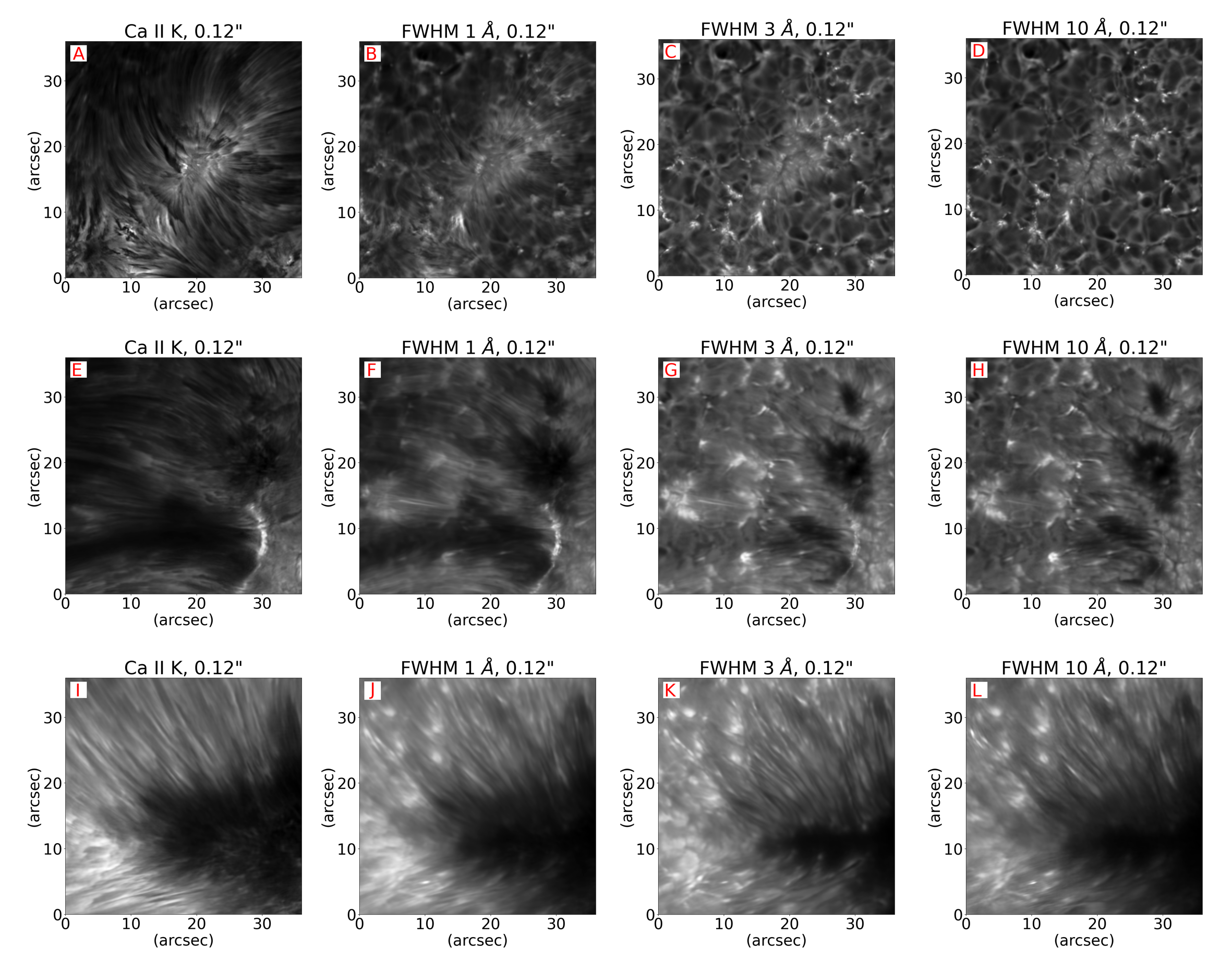}
 }
	\caption{Examples of the observed (left column panels) and spectrally degraded (all other panels) images at the \ca  line core of the three studied solar regions, namely a  quiet Sun area (top row panels), a region with plages and several pores (middle row panels), and a  sunspot with umbra and penumbra (bottom row panels), all observed close to disc center. For each target, from left to right we show original CHROMIS observations  and data degraded with Gaussian kernels with FWHM of 1, 3, and 10~\AA. Each  image is  shown using the  intensity interval that enhances the visibility of the solar features therein.  }
	\label{fig5}
\end{figure*}

Figure \ref{fig2} shows examples of the observations analysed in our study. In particular, we show the data taken at the \fe 6302~\AA \ line continuum (panels A--C), in the \ca line core (panels D--F), at roughly the secondary minimum K$_\mathrm{1V}$ in the violet wing of the \ca line (panels G--I), and in the red wing of the line at  
+1.05~\AA \ from the line core (panels J--L), for the QS (left column panels), NA (middle column panels), and SP (right column panels) regions. We show the observations of the same solar features at the above diverse positions along the \ca line in order to highlight their different appearances as is in the case of the data of some existing series of \ca solar observations, e.g., the set of the Meudon Observatory that includes data  at the centre and blue wing of the \ca line. 

The photospheric \fe data of the three observed regions show the regular granulation pattern in QS areas (panel A), elongated granules around the pores (panels B) and other larger scale features, such as the umbra and penumbra (panel C) in the SP target including bright umbral dots and dark and bright filaments, respectively. The \ca line core images display the chromosphere above the same areas, consisting of thin bright and dark fibrils everywhere, except in the umbra and inner part of the penumbra, and of several localised bright features (panels D--F).
On the other hand, the data taken at the K$_\mathrm{1V}$ and in the red wing of the \ca line display disc features typical of the upper photosphere and temperature minimum, namely the reversed granulation pattern and the bright features in the  intergranular lanes that are due to magnetic field concentrations and/or acoustic grains (panels G, J). These patterns are more evident in the red wing observations.   
Umbral regions are dark in all \ca observations, while in line core and line wing data penumbral areas are only partially so (panels F, I, L). We note extended brightenings may also appear in the umbral regions (in the chromosphere) due to umbral flashes. The pores are dark in  line wing observations (panels H--K), while line core data display a small-scale (transient) brightening at higher atmospheric heights above them (panel E). All the \ca images show brightness patterns due to granulation and associated with dark magnetic features. Most of the bright/dark fibrils in the core images seem to overlie the bright patches displayed by the wing images (panels D, G, J and panels E, H, K). The fibrils observed in the \ca line core data  
are only slightly curved. 

It is worth noting that the diverse solar regions described above are characterized by rather different \ca line profiles and are only representatives of the regions sampled in these observations that could look different based on, e.g., other magnetic topology. In Fig.~\ref{fig3} we compare \ca line profiles obtained from spatial averaging of the \ca line data in quiet  areas of the  QS, NA, and SP regions, overplotted on the high-resolution spectrum from the atlas of \citet{Delbouille1973}. Similarly to the latter, all the profiles derived from the observations show the K$_\mathrm{1V}$ and K$_\mathrm{1R}$ secondary minima in the violet and red part of the spectrum, respectively, the reversed line centre K$_{3}$, and the two K$_\mathrm{2V}$ and K$_\mathrm{2R}$ peaks, with K$_\mathrm{2V}$ stronger than K$_\mathrm{2R}$ as reported in the literature. The above line features are almost inappreciable on the top right panel of the figure resulted from the QS observations, but they increase significantly in the data from SP regions (bottom right panel). On the other hand, in the NA case the  K$_{3}$ line centre, and the two K$_\mathrm{2V}$ and K$_\mathrm{2R}$ peaks are not visible (middle right panel).

We notice that the separation of the line peaks and their width in all the QS spatially-averaged observed profiles are close to the ones in the atlas profile. However, this is not the case for profiles extracted from individual pixels in the analysed observations. Figure \ref{fig4} displays some examples of these profiles, from the 20~$\times$~20 pixels wide areas marked with numbers and coloured boxes in the QS, NA, and SP observations of Fig.~\ref{fig2}.   For each area under investigation, we also show the line profile (black solid line) derived from averaging the data available at each spectral position of our spatially resolved  observations. 

The profiles in Fig.~\ref{fig4} show different characteristics: there are profiles with emission peaks significantly increased (QM box 1, top left panel) with respect to the ones in Fig.~\ref{fig3}, as well as profiles with fuzzy K$_\mathrm{2V}$ and K$_\mathrm{2R}$ peaks and a reversed strength than reported in the literature (QG box 2, top right panel), and profiles lacking the K$_3$ minimum (PO box 3, middle left panel; PE box 5, bottom left panel; and UM box 6, bottom right panel).  

We note that the profiles in Fig.~\ref{fig4} derive from the areas randomly selected to represent diverse solar features in the ambient to which they belong. Indeed, analysing slightly different areas in their surroundings, the obtained profiles often retain the characteristics of the individual profiles in Fig.~\ref{fig4}, but this is not always the case. Besides, the mean profiles derived from other boxes nearby the ones in Fig.~\ref{fig2} can slightly differ from the average profiles in Fig.~\ref{fig4}. This is particularly evident in results derived from QM and QG data. Some examples for these data are given in Appendix A.

\subsection{Effect of spectral bandwidth}

We wondered how the \ca observations and measured line profiles presented in the previous Section are affected by the characteristics of the instruments employed for \ca measurements, in particular their spectral bandwidth and spatial resolution. 
To answer the above question, in Fig.~\ref{fig5} we show an about 25$\times$25\arcsec \ FoV of the original CHROMIS observations at the \ca line centre for the three analysed solar regions (panels A, E, I) and corresponding data after their spectral degradation with  kernels having FWHM of 1, 3, 10~\AA \ (all other panels). The regions shown in Fig.~\ref{fig5} are the ones marked with red boxes in Fig.~\ref{fig2}.  

As expected, smearing the original data with a spectral kernel leads to mixing of photospheric and chromospheric emissions. The fibrils filling the original QS FoV (panel A) are faintly detectable with a spectral degradation of 0.6~\AA \ (not shown) and almost   
no longer seen with 1~\AA \ 
bandwidth (panel B), which allows PO and UM regions to manifest themselves with spatial scales and intensities (panels F, J) close to the ones displayed by the same features in  images acquired with a spectral degradation of 3--10~\AA \ (panels G, K and panels H, L). Observations of the QS region 
with such bandwidths  show the reversed granulation  and dot-like bright features (panels C--D) seen in the original observations at the red wing of the \ca line (shown in Fig. \ref{fig2}, panel J). 

Among the three regions studied here, the NA target is the one visually showing the largest change in its appearance with the various spectral degradation kernels. This applies, in particular, to data obtained with bandwidths up to 1~\AA \ and those derived from bandwidths in the range 
[3,10]~\AA. Indeed, the former data show fibrils similar to penumbral filaments, while in the latter data bright granular elements appear around the pores. These granular  features, which seem to be raised above the pores, are outlined by thin dark boundaries, as reported in e.g. \citet{lites2004} from a study of SST photospheric data. The same features also resemble the patterns attributed to sea-serpent magnetic configurations in e.g. \citet{murabito2021}. 
We also notice that the panels in Fig.~\ref{fig5}, illustrating spectrally degraded data, show brightness patterns with similar position and extension in QS   observations (panels B--D), and partly so in the SP data (panels J--L). However, the average \ca brightening of disc features in the above degraded images is always lower than in original CHROMIS observations. We also note that degrading the observations with larger kernels leads to appearance of smaller scale structures and finer details (panels D, H, L), and reversed granulation pattern (panels C--D).   

\begin{figure}
\centering{
\includegraphics[scale=0.98, trim=10 10 0 0]{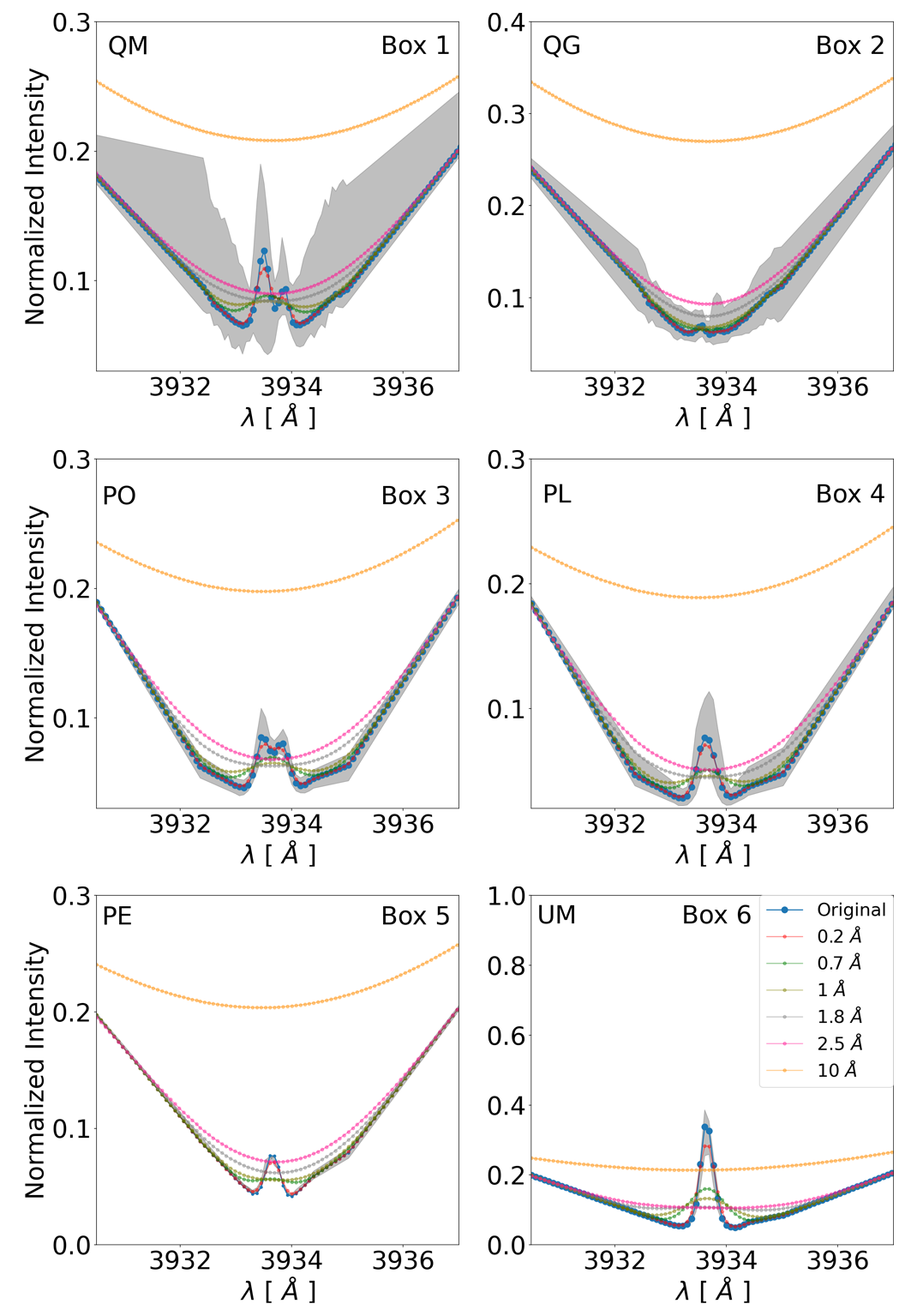}}
\caption{\ca line profiles derived from data averaging over the areas marked with numbers and coloured boxes in Fig.~\ref{fig2}. The diverse profiles are color-coded depending on the degradation applied to the original data and according to the legend in the bottom-right panel. From top to bottom we show the profiles for the selected areas in the QS (top row panels), NA (middle row panels), and SP (bottom row panels) observations, respectively. The profiles are normalized to the continuum intensity at 4000~\AA \ (not shown). The grey shaded areas cover the range of values measured in the analysed regions at each spectral position.}
	\label{fig6}
\end{figure}

The above effect of spectral degradation of the data is even more evident when considering the \ca line profiles. In Fig.~\ref{fig6} we show line profiles extracted from the original and spectrally degraded observations of the various targets in Fig.~\ref{fig5}. We report spatially averaged profiles computed over the regions marked with numbers and coloured boxes in Fig.~\ref{fig2}.

Figure \ref{fig6} shows that a spectral degradation 
of the order of 10~\AA \ results in an increase of the K$_3$ intensity of the \ca line core, and of the intensity all along the line profile, in all the areas except in the UM one. On the other hand, a spectral degradation of only 2.5~\AA \ results in an increase of the K$_3$ intensity  in QM  and QG areas, being more evident in the latter case, and a decrease of the same quantity in all the other targets. Noteworthy, spectral bandwidths from 0.7~\AA \ to 2.5~\AA \ have similar effects on the line profiles derived from the PO and PL areas.

Figure \ref{fig6} also shows that the features of the \ca line typical of each studied areas disappear completely in observations characterized by a bandwidth  
of 1.8~\AA. It is worth noting that this spectral bandwidth is lower  
than the ones used by most sites currently performing observations in the Ca II K line with optical filters, but higher than all the ones of spectroheliograph data \citep[see][]{chatzistergos2022fass}. 

We note that the various panels in Fig.~\ref{fig6} also include shaded areas, which cover the range of values measured over each analysed region. These shaded areas manifest the large heterogeneity of the \ca line profiles in the studied  observations when considered at the  spatial  resolution of the CHROMIS observations. This heterogeneity is particularly evident in the QS data, that seem to be especially affected by intensity oscillations  due to hydrodynamic pressure modes \citep[p-modes,][]{Leighton1960}, and very noticeable in the line wings, where small changes in wavelength due to p-modes lead to large variations in intensity. On the other hand, intensity fluctuations are also evident in the profiles derived from the NA areas, while they are not appreciable in the profiles obtained from the SP field. The intensity fluctuations in both these regions are most likely attributable to  magneto-hydro-dynamical oscillations \citep[MHD-modes, e.g.][]{spruit1982} and to small scale transient brightening from magnetic reconnection events. 

Overall, the results presented above suggest that the effect of spectral bandwidth on \ca line profiles depends on both the employed bandwidth and observed solar region.

\subsection{Effect of spatial resolution}

\begin{figure*}
\centering{
\includegraphics[scale=0.87]{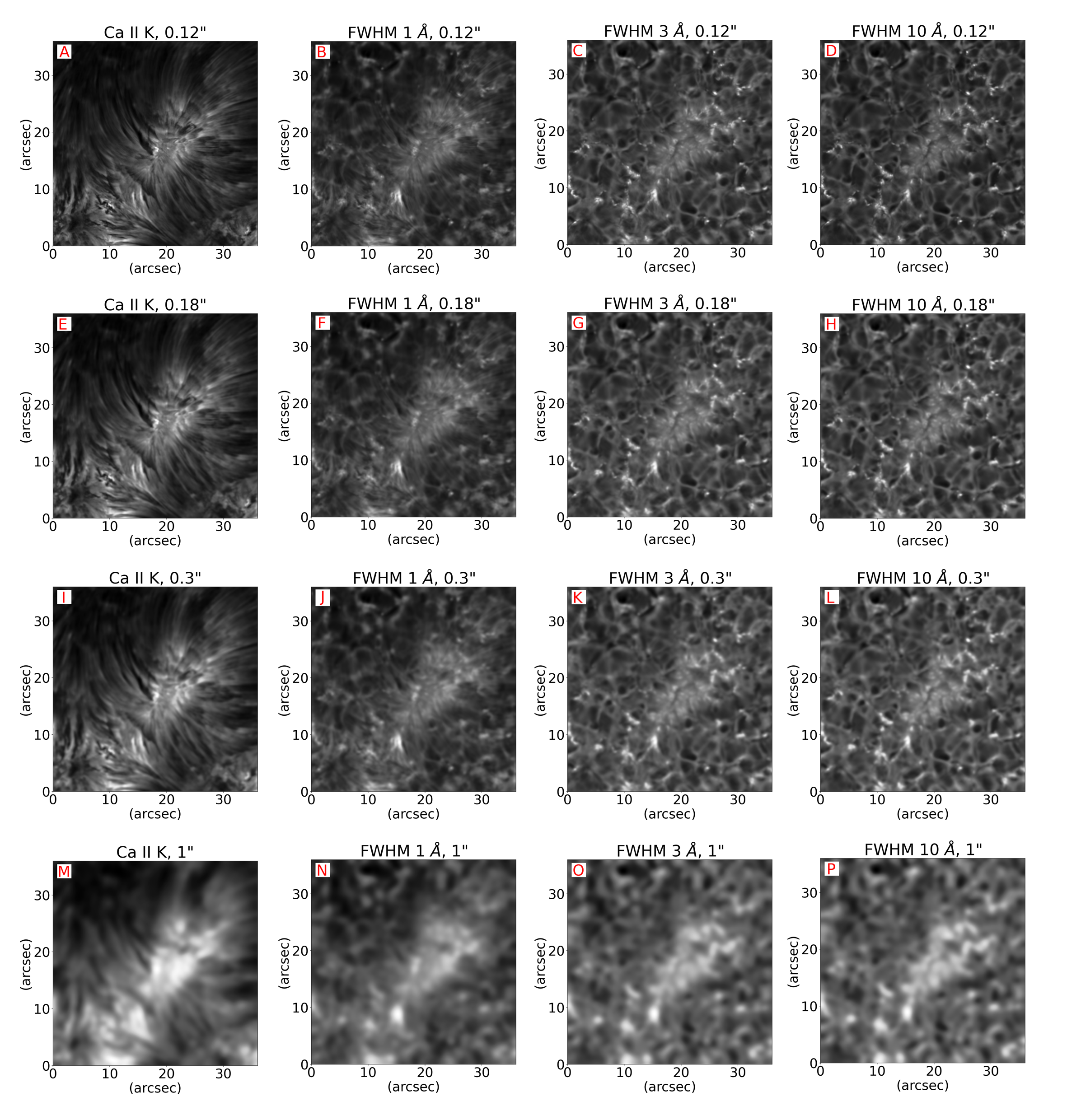}
}
\caption{Examples of the original  (panel A) and degraded (all other panels) images at the \ca  line core of the QS region, to account for the  diverse bandwidths and spatial resolutions of the most prominent series of available \ca observations. 
Each row shows examples of data characterized by a given pixel scale and by different bandwidths. From top to bottom, we show data at the original pixel scale of the CHROMIS  observations (panels A--D) and degraded to a spatial resolution  of 
0.18\arcsec  \ (panels E--H), 0.3\arcsec  \ (panels I--L), and 1.0\arcsec  \ (panels M--P), as is in the case of the {\sc Sunrise}/IMaX, Hinode/SOT, and SDO/HMI observations, respectively. For each of these observations, from left to right we show the data at the spectral resolution of the CHROMIS  observations of 0.12~\AA, and spectrally degraded with Gaussian kernels with FWHM of 1, 3, and 10~\AA. Each  image is  shown using the  intensity interval that enhances the visibility of the solar features therein. }
	\label{fig7}
	\end{figure*}

We then considered the impact of spatial degradation on our data. As an example, in Fig.~\ref{fig7} we show data derived from spatial degradation of the original observations of the QS region, under the diverse spectral degradations analysed above. For the sake of comparison, we also show the original CHROMIS observations, characterized by a spatial resolution of 0.078\arcsec. More examples for the QS, NA, and SP regions 
are given in Appendix B and presented  in the following. For all these examples, we considered the spatial resolution of the {\sc Sunrise}/IMaX, Hinode/SOT, and SDO/HMI observations acquired with a pixel scale of 0.09, 0.15, and 0.5\arcsec/pixel, respectively.  Moreover, we considered the case of the synoptic full-disc observations characterized by a moderate pixel scale of 2\arcsec/pixel, as in the data acquired at e.g. the Rome Observatory.  Results for the latter case are also reported in Appendix B. 

For the data representative of all the above pixel scales, we considered spectral degradations resulting from the application of  Gaussian kernels with FWHM of 1, 3, and 10~\AA \ to the original CHROMIS data characterised by a 0.12~\AA \ spectral resolution. The above FWHM values are distinctive of the bandwidth of optical filters employed for modern observations in the \ca line, at e.g. the Rome, Kanzelh\"ohe, and San Fernando Observatories, respectively. 

From the data in Fig.~\ref{fig7}, we notice that the spatial degradation impacts almost equally the data characterised by a spatial resolution better than 0.3\arcsec. Besides, we observe that the appearance of the data is barely affected by employed spectral bandwidths  in the range [1,10]~\AA. This also applies to data degraded to the 4\arcsec \ spatial resolution typical of full-disc observations, which show the main features of the observed regions seen in the higher resolution data preserved on larger spatial scales, and  less details at the small scales.   
These findings are in agreement with the results reported by \citet{Chatzistergos2021} concerning reconstructions of irradiance variations and unsigned magnetograms derived from \ca observations, which both resulted to be only slightly sensitive to the bandwidth 
of the analysed observations in the range  [0.09,9]~\AA \ and to the data spatial resolution in the range of $\sim$[2,11]\arcsec .   

As mentioned above, more examples of solar regions data at the original resolution of the CHROMIS observations and derived from the various degradation kernels applied to them are given in Appendix B. In particular, we show  examples of the QS region observed in the red wing of the \ca line at +1.05~\AA \ from line core and of the NA and SP areas observed at the \ca line core. From these additional examples we note that regardless of the spatial and spectral degradation applied to the data, the QS target shows different characteristics when observed in the \ca line core and in the red wing of the \ca line, but the latter resemble observations at the K$_\mathrm{1V}$.  On the other hand, we notice that the granular bright features surrounding  the pores in the NA region observed at the \ca line core with the 0.078\arcsec \  original spatial resolution of the CHROMIS data, and with bandwidths in the range [3,10]~\AA, are retained in the data degraded to a spatial resolution up to 0.3\arcsec, while they form unresolved larger scale bright features in the data characterised by a spatial resolution of 1\arcsec. We also note that the arc-shaped feature with largest intensity gradient seen in the original NA observations is visible in all the data derived from spatial  degradation up to 1\arcsec \ and spectral bandwidths  up to 3~\AA, although with lower intensity. However, in addition to this arc-shaped feature, spatial and spectral degradations bring out in the observed field several dot-like and small-scale bright structures unseen in the original data. This also applies to observations of the SP target at the \ca line core. In these data we notice that the appearance of the umbral and penumbral regions is only barely affected by a spatial degradation up to 1\arcsec \ and  spectral degradation up to 1~\AA. Indeed, the umbral and penumbral areas are almost unaffected by the spatial degradation applied to the data, while their extension slightly decreases in the data degraded with spectral kernels larger than 1~\AA.

In Appendix B we also show examples of the QS, NA, and SP regions observed at the \ca line core and reported in Fig.~\ref{fig2}, as returned when degraded to a spatial resolution of 4\arcsec \ and to spectral bandwidths in the range [0.12,10]~\AA. We notice that these data show unresolved bright patterns that are rather unaffected by the spectral degradation applied to the data. We also report a marginal decrease of the umbral area for data with spectral degradation larger than 1~\AA. It is worth noting that the umbral region in the degraded data shows almost same extension than the same region in the full resolution photospheric CRISP observations and  chromospheric CHROMIS data at the red wing of the \ca line.

\subsection{Focus on line parameters}

\begin{figure*}
\centering{
\includegraphics[scale=0.90, trim=10 10 0 0]{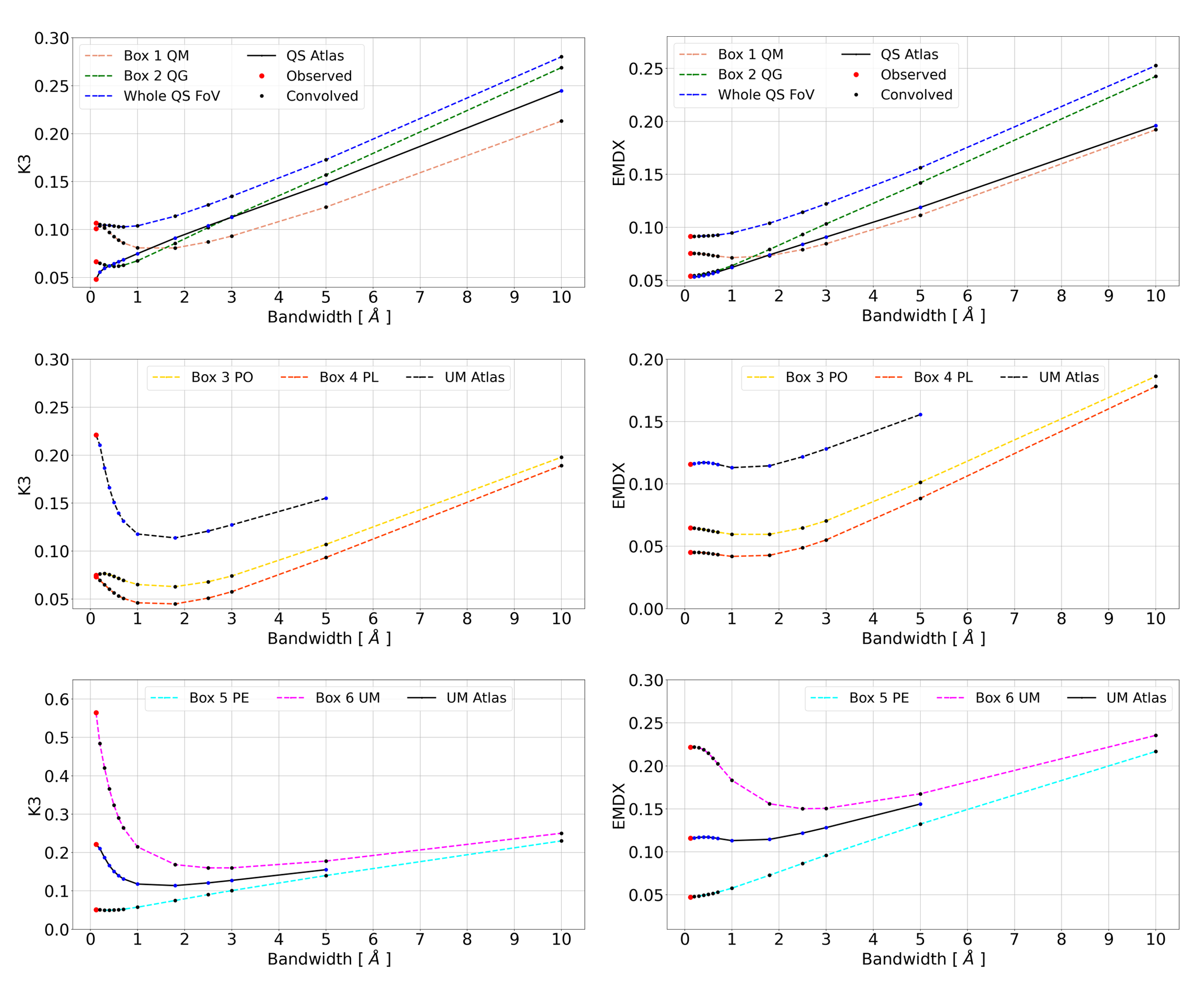}

}
	\caption{Dependence of the K$_3$ (left column panels) and $EMDX$ (right column  panels) parameters estimated from the \ca line measurements on the QS region and on the QM and QG areas therein (top row panels), PO and PL areas in the NA target (middle row panels), and PE and UM areas selected in the SP observation (bottom row panel), depending on the spectral degradation applied to the analysed data. Overplotted to each panel is the variation of the  parameter derived from the relevant atlas data (black solid line) after their spectral degradation with the various kernels used in our study. Find information about the atlas data in Sect. 2.2. 
	\label{fig8}}
\end{figure*}

 Figure \ref{fig8} describes the impact of the spectral degradation on the K$_3$ and $EMDX$ parameters derived   
from the diverse solar data considered in our study. In particular, we display the variation of the K$_3$ (left  panels) and $EMDX$ (right panels) parameters depending on the spectral kernel applied to the line profiles averaged over the whole QS region and over the QM (box 1) and QG (box 2) areas selected therein (top panels), as well as over the PO (box 3) and PL (box 4) areas in the NA region (middle panels), and the PE (box 5) and UM (box 6) areas in the SP target (bottom panels). We also report the dependence of the same parameters estimated by using the respective reference atlas data. 

Figure \ref{fig8} (top panels) shows that the K$_3$ and $EMDX$ parameters derived from the QS and QG (QM) degraded data increase gradually for  
spectral bandwidths larger than 1~\AA \ (2~\AA) and are almost unaltered in observations with bandwidths in the range [0.12,1.0]~\AA \ ([0.12,2.0]~\AA). Therefore the sensitivity to spectral degradation of both parameters obtained from QS and QG areas is slightly different from the one derived from the QM region. We notice that the trend  derived from the whole QS target reproduces the variation of the parameters derived from atlas measurements better than obtained from the other regions, except for a difference in the estimated values. The increase of the K$_3$ ($EMDX$) values is of the order of about 2~\%/\AA \ (1.5~\%/\AA) for the QS observations with spectral bandwidth  larger than 1~\AA. 

We also studied the K$_3$ and $EMDX$  parameters in the profiles derived from the PO (box 3), PL (box 4), PE (box 5), and UM (box 6) observations. As shown in Fig.~\ref{fig8} (middle panels) the variation of the K$_3$ and $EMDX$ parameters derived from the PO and PL areas are close to each other for spectral bandwidth of about 0.2~\AA \ and exhibit a similar behaviour for all larger bandwidths, with differences that tend to decrease for bandwidths larger than 1~\AA. We also note that for spectral bandwidths in the range [0.12,10]~\AA, the results derived from the NA region (middle panels) are rather similar to those obtained from the QM area (top panels). The same applies to the findings from  the PE region (bottom panels), but limited to spectral degradations in the range [1,10]~\AA.   
Besides, Fig.~\ref{fig8} (bottom panels) displays that for UM regions the values of the K$_3$ and $EMDX$ parameters estimated from the spectrally degraded data increase gradually for bandwidths larger than about 3~\AA, while the values of both parameters decrease in observations with a spectral degradation in the range [0.12,2]~\AA. This decrease of the values of the K$_3$ and $EMDX$ parameters is of the order of 35~\%/\AA \ and 18~\%/\AA, respectively, while the increase of the parameters for the data with bandwidths in the range [3,10]~\AA \ is of the order of 1~\%/\AA. Similarly to results from the QS and QG regions, the parameters computed for the PE area are almost unaltered in observations with spectral bandwidth in the range [0.2,1.0]~\AA, while they increase gradually of about 1~\%/\AA \ for the data with bandwidths in the range [1,10]~\AA. 

We notice that the K$_3$ and $EMDX$ parameters computed on each observed region vary with the spectral degradation applied to the data in a rather similar way. Therefore, in the following we further consider the characteristics of the K$_3$ variation only.  

We report that the best relation describing the changes of the K$_3$ parameter on the spectral bandwidth of the data is a polynomial function of the fifth order that is represented by the equation 
\begin{equation}
K_3=ax^5+bx^4+cx^3+dx^2+ex+f, 
\end{equation}
where $x$ is the spectral bandwidth   considered, expressed in \AA. Table \ref{t1} summarizes the $a-f$ parameters for the various studied data. 

We find that the fitted curves follow the data derived from the observations with minute deviations. Indeed, all the fits give  
mean standard deviation between original and fitted curves in the range [1.8,15.4]$\times 10^{-4}$, except for the UM data, which show a mean deviation of about 62$\times 10^{-4}$.   
However, we notice that the fitting of the UM data is slightly better represented by a polynomial function of the sixth order\footnote{Described by the equation K$_3=ax^6+bx^5+cx^4+dx^3+ex^2+fx+g$, with  the following coefficients:   
a=(0.8$\pm$0.2)$10^{-5}$,   
b=(-0.02$\pm$0.3)$10^{-4}$,  
c=(0.01$\pm$0.2)$10^{-3}$, 
d=(-0.5$\pm$7)$10^{-2}$, 
e=(0.01$\pm$0.0009)$10^{-2}$,         
f=(-0.1$\pm$0.05)$10^{-1}$, 
g=(6.8$\pm$0.0008)$10^{-1}$.}, which results in an average deviation of about  26$\times 10^{-4}$.  Note that we derived the mean deviation from the average difference between original and fitted values, and the 1$\sigma$ uncertainty from the diagonal values of the covariance matrix of the optimal coefficients computed with the \textit{scipy.optimize.curve$\_$fit} function in the \textit{Python} language.   

\begin{figure}
\centering
{
\includegraphics[scale=0.18,trim=90 50 100 0]{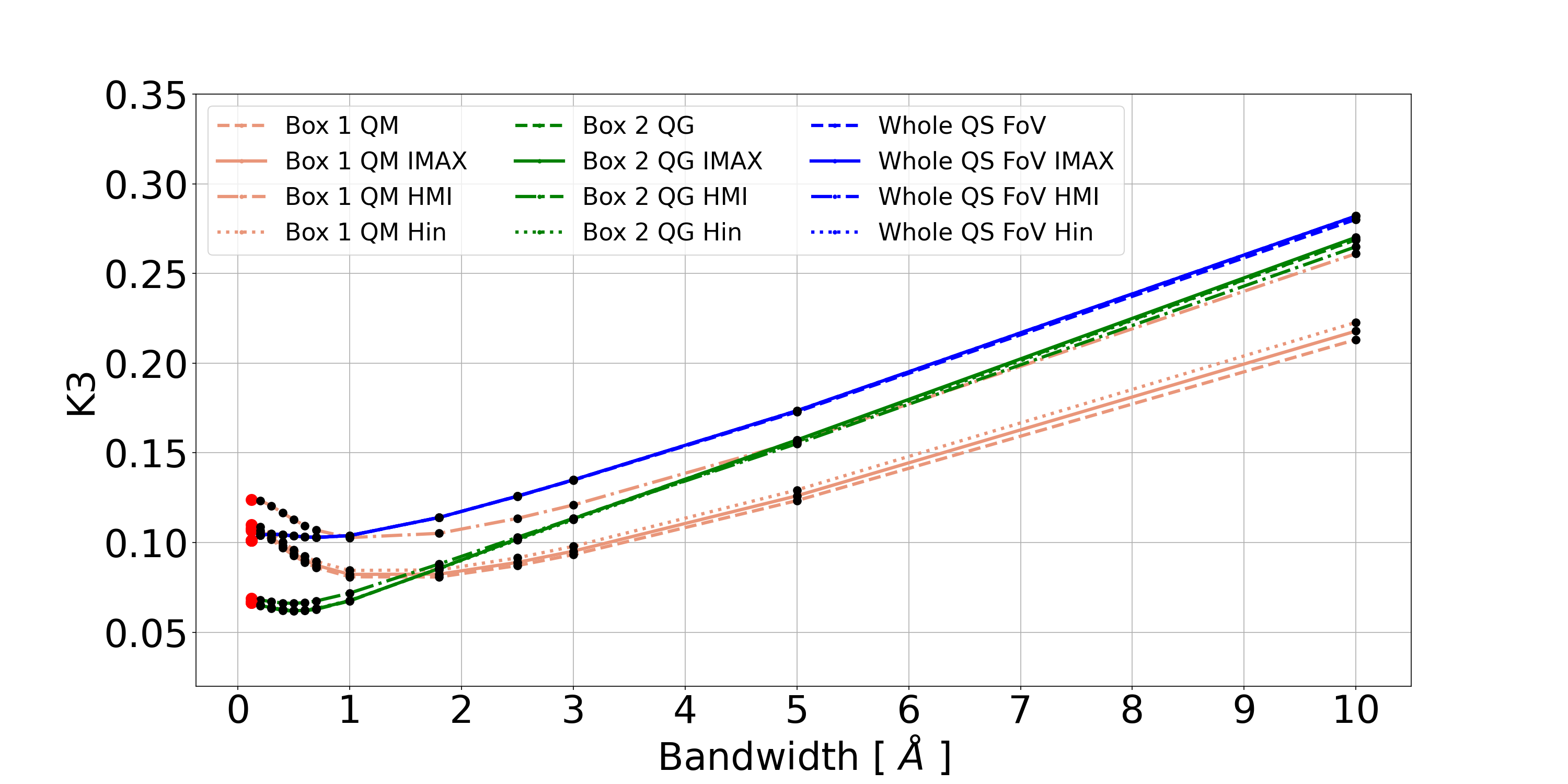}
\includegraphics[scale=0.18,trim=90 0 100 0]{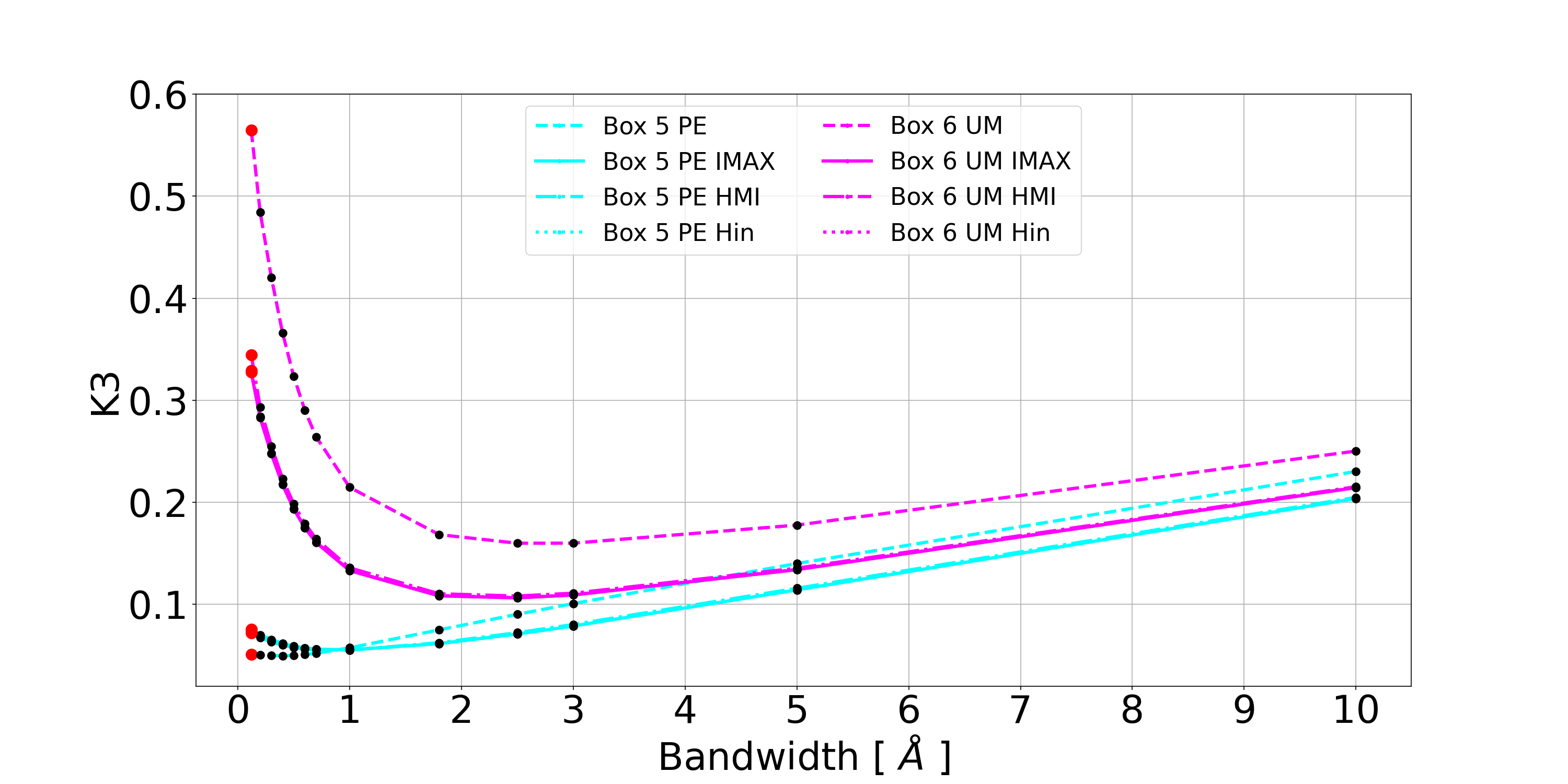}

}
	\caption{Dependence of the K$_3$ parameter estimated from the \ca line measurements on the QS, QG, and QM areas (top panel) and on the PE and UM regions (bottom panel), depending on the spectral and spatial degradation applied to the analysed data, representative of various observations. Find more details in Sects. 3.2 and 3.3.}
	\label{fig9}
\end{figure}

We then considered the effect of the spatial degradation on the values of the K$_3$ parameter estimated for the various solar regions and spectral 
bandwidths considered in our study. 

Figure \ref{fig9} displays the variation of the K$_3$ parameter derived from the line profiles averaged over the whole QS region and selected QM and QG areas (top panel) therein, as well as over the PE and UM regions (bottom panel)  of the SP target, by considering the CHROMIS data and those obtained from the various spectral and spatial kernels applied to them. Similarly to findings in Fig.~\ref{fig8}, results from the PL and PO areas are close to the ones reported for the QM and UM regions, respectively, and are thus not shown. 

Results in Fig.~\ref{fig9} suggest that the effect of the spatial degradation on K$_3$ vales depends on the solar target. It is significant for the selected areas representative of the QM region (box 1) in the QS target, and of the PE (box 5) and UM (box 6) areas in the SP region. In particular, for observations of the QM, PE, and UM regions, 
K$_3$ changes of about  [2,30]~\%, -[5, 20]~\%, and -[15,40]~\% maximum  
when the data are degraded to  observations with spatial resolution from 0.18\arcsec \ to 1\arcsec, which represent the cases  of the {\sc Sunrise}/IMaX and SDO/HMI observations, respectively. 
K$_3$ changes in QM, PE, and UM areas from about 5~\%/\AA \ to 20~\%/\AA,  15~\%/\AA \ to 30~\%/\AA, 0.1~\%/\AA \ to 30~\%/\AA \ minimum to maximum  when considering data with a  spectral degradation given by bandwidths in the range [2,10]~\AA.  
For the QG and QS areas the spatial degradation affects only minutely the estimated value of the K$_3$ parameter.


\begin{table*} \begin{center}
\caption{Results from K$_3$ parameter fitting.\label{t1}}
\begin{tabular}{cccccccc}
\hline
\hline
Region & a & b & c  & d  & e  & f &mean SD\\
 &  [$10^{-5}$]  &  [$10^{-4}$] &  [$10^{-3}$] &  [$10^{-2}$] &  [$10^{-2}$] & [$10^{-1}$] & [$10^{-4}$]\\
\hline
QM             &-3.5$\pm$2.5      & 8.8$\pm$4.9     & -8.1$\pm$2.9   & 3.5$\pm$0.7      & -6.2$\pm$0.6    & 1.1$\pm$0.01 & 6.8\\
QG             &-9.5$\pm$0.76     & 19$\pm$1.4      & -14 $\pm$0.8   & 4.3$\pm$0.2      & -3.4$\pm$0.2    & 0.7$\pm$0.004 & 2.0 \\
QS          & 2.4$\pm$2.9      & -4$\pm$5.5      &  1.5$\pm$3.3   & 2.8$\pm$0.7      & -0.2$\pm$0.7    & 1$\pm$0.02  & 8.1\\
PE             &-7.2$\pm$0.7      & 15$\pm$1.3      & -10$\pm$0.8    & 3.1$\pm$0.2      & -1.8$\pm$0.2    & 0.5$\pm$0.004  & 1.8\\
UM             &-141$\pm$22        & 294$\pm$42      & -209$\pm$25    & 65$\pm$5.8       & -92$\pm$5       & 6.5$\pm$0.1  & 61.5\\
QS Atlas         &5.7$\pm$3.5       & -11$\pm$6.5     & 7.8$\pm$3.9    & -2.2$\pm$0.9     & 4.6$\pm$0.8     & 4.6$\pm$0.2  & 9.2\\
UM Atlas         &-136$\pm$80       & 191.4$\pm$86.7  & -102.9$\pm$31.7& 26.7$\pm$4.8     & -32.7$\pm$2.8   & 2.6$\pm$0.05   & 15.4\\
\hline
\end{tabular}
\normalsize 
\tablecomments{Columns are: solar region over which the relationship was studied, best fit
parameters (a, b, c, d, e, f) with their 1$\sigma$ uncertainties, and the mean standard deviation (SD) between original and curves of the fits. Find more details in Sect. 2.2 and Sect. 3.3.}
\end{center} \end{table*}

\section{Discussion}
\label{sec4}

\begin{figure}
\centering
{
\includegraphics[scale=0.54,trim=0 30 0 0]{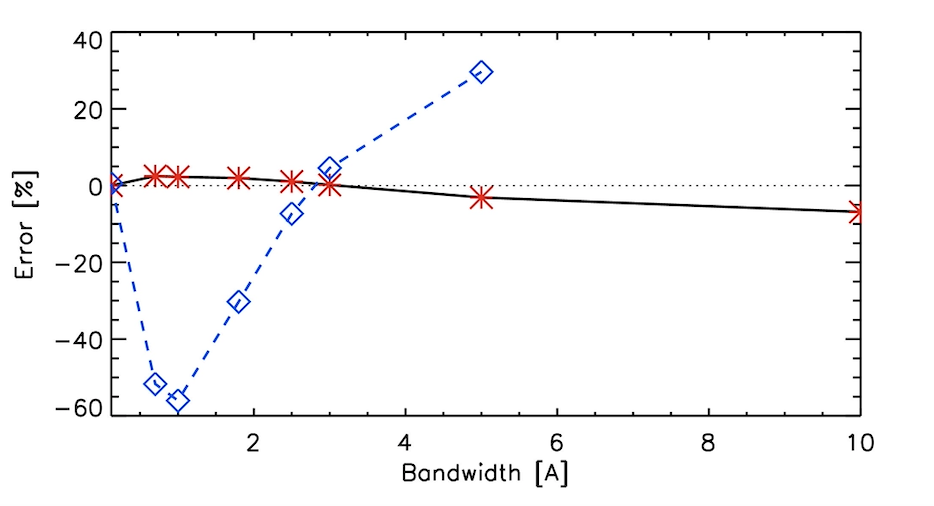}
}
	\caption{Error [\%] in the K$_3$ values estimated from atlas data representative of QS region (solid black line, red star symbols), and NA and SP (dashed blue line, diamond blue symbols) regions, depending on the spectral degradation applied to the data. For the sake of clarity overplotted is the null error line (dotted line). 
	}
	\label{fig10}
\end{figure}

\begin{figure}
\centering
{
\includegraphics[scale=0.18,trim=90 50 100 0]{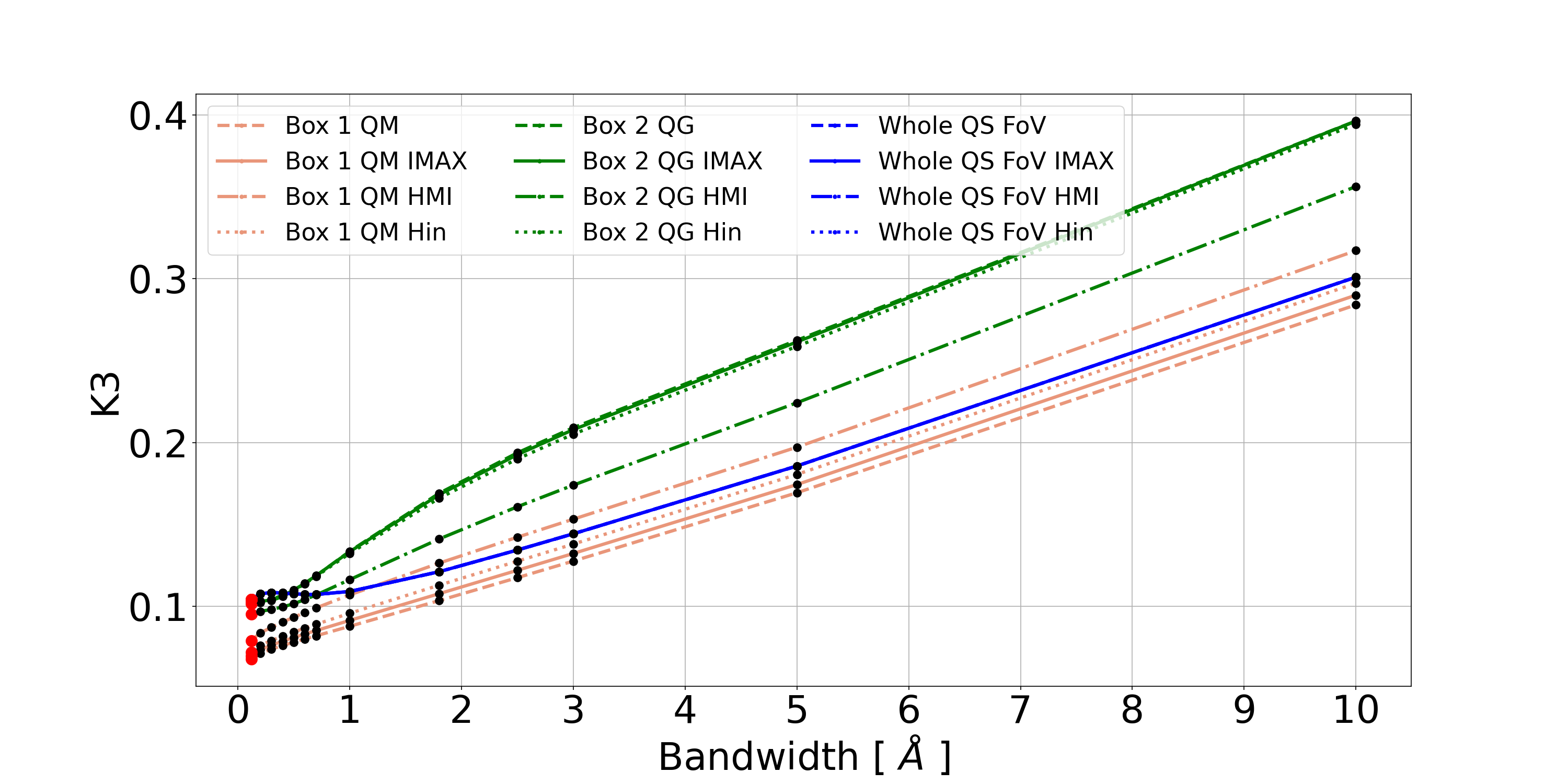}\\
\includegraphics[scale=0.18,trim=90 0 100 0]{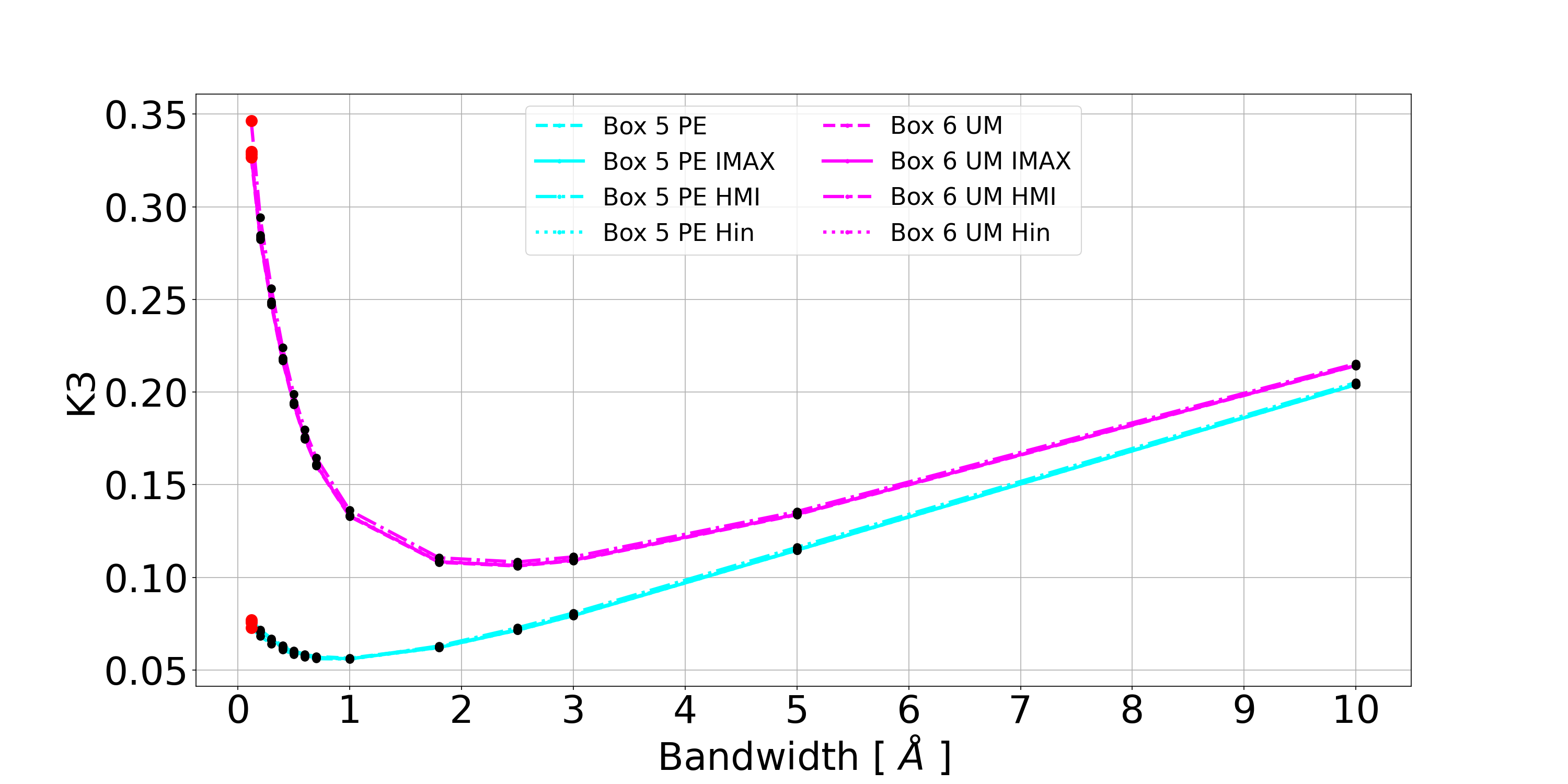}

}
	\caption{Dependence of the K$_3$ parameter estimated from the \ca line measurements on the QS, QG, and QM areas (top panel) and on the PE and UM regions (bottom panel), depending on the spectral and spatial degradation applied to the analysed data, representative of various observations. Here the reported values derive from average \ca line profiles computed pixelwise on spectrally and spatially degraded data. Find more details in Sect. 4.}
	\label{fig9b}
\end{figure}

\begin{figure}
\centering
{
\includegraphics[scale=0.2,trim=90 50 100 0]{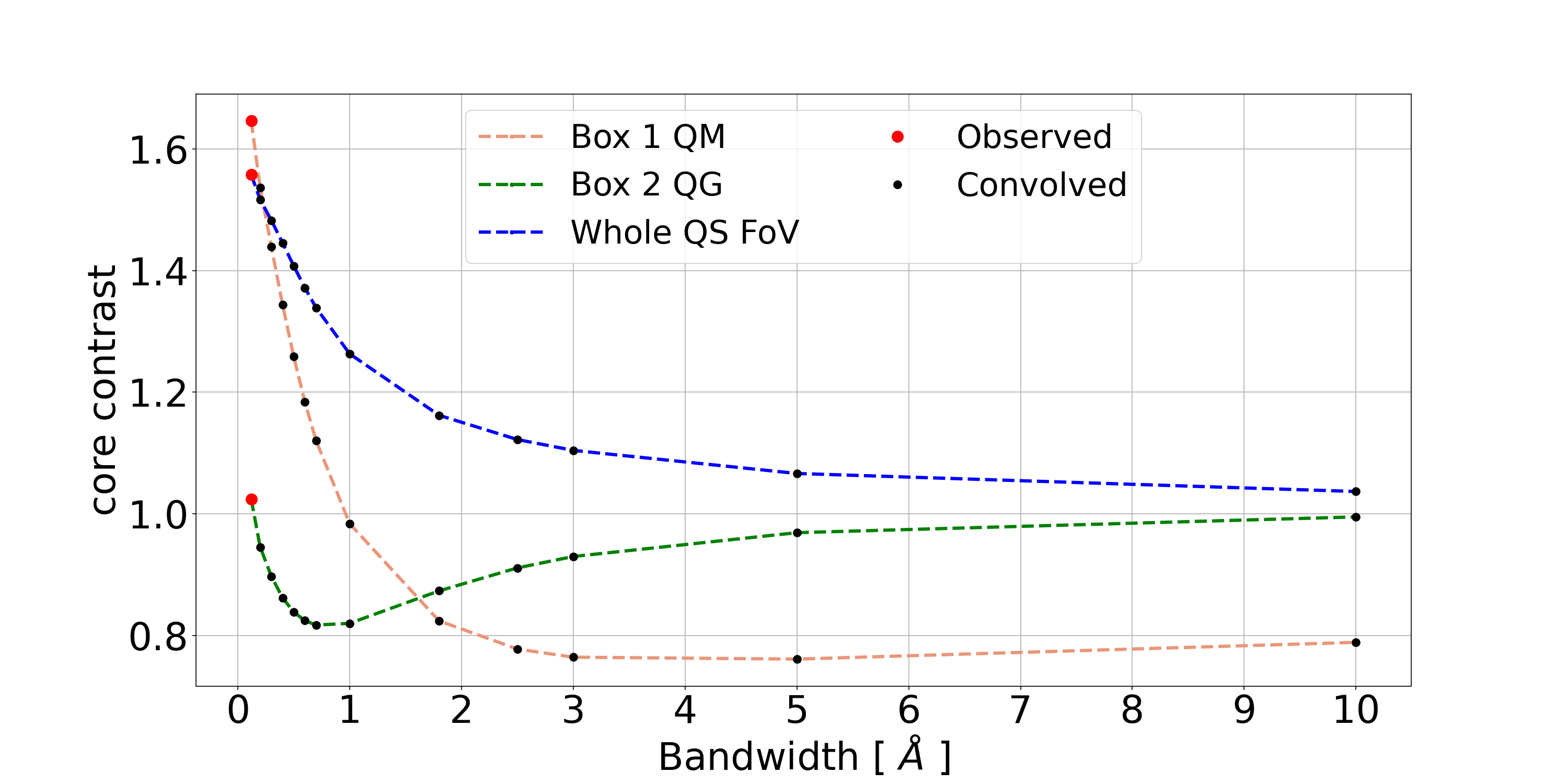}
\includegraphics[scale=0.2,trim=90 50 100 0]{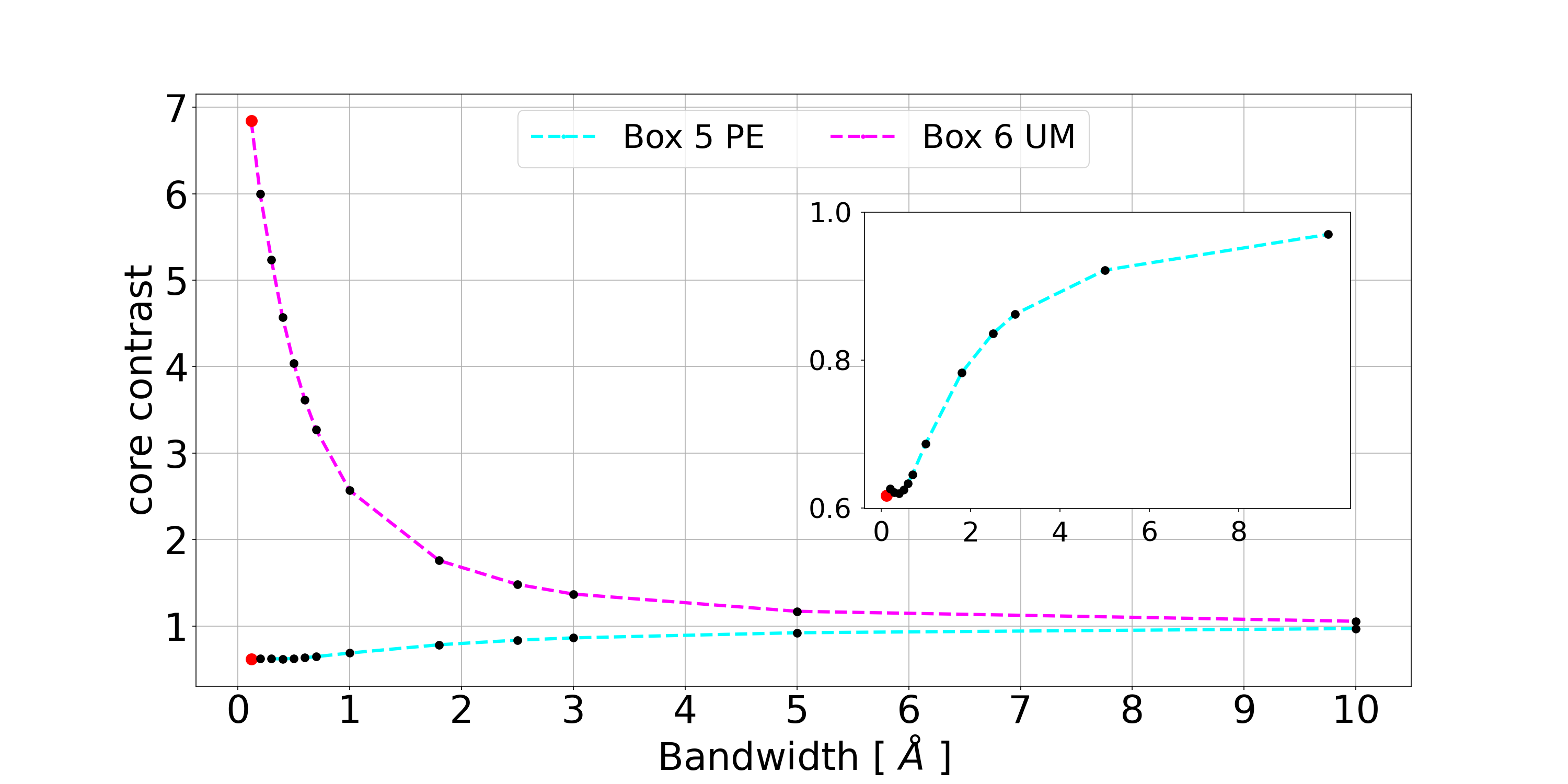}
}
	\caption{Dependence of the estimated line core intensity contrast on the spectral resolution of the analysed data, on the whole QS region and on the QM and QG areas therein (left panel), as well as on the PE and UM areas selected in the SP observations (right panel). See Sect. 4 for more details.}
	\label{fig11}
\end{figure}

The observations analysed in our study were compensated for residual effects of atmospheric turbulence, and calibrated with atlas measurements before their degradation to match the diverse spectral and spatial resolutions considered. However, they are not compensated for the stray-light contamination, which is estimated to be low and mostly to come from the SST instruments \citep{scharmer2019}. Intensity values analysed in our study may thus suffer from incomplete compensation of stray-light, because the spectral intensity calibration applied to the data using the disc center atlas spectrum as reference does not remove all instrumental effects. Based on comparison between intensity values measured in the QS target and in the atlas reference we expect  compensation for stray-light to affect line core intensities reported from our study with an increase of their values of about 40~\% and 80~\% in QS and PO penumbral data, respectively. Moreover, we would like to emphasize that the observations and line intensities reported in this study might not be representative of all the similar solar and stellar data. Indeed, we highlight that they refer to the specific level of solar magnetism and solar activity framed by the observations analysed in our study. 

In addition, we have reported results based on line profile data that combine measured values in the inner part of the line and extrapolated values in the line wings. As stated above, these data simulate low resolution spectral observations in the line wings and high resolution observations in the line core. In addition, the extrapolated values do not account for the many lines that populate the spectra adjacent to the \ca line core. We investigated the impact of the unaccounted line blending in the obtained results. To this end, we analysed atlas data and applied to these all the processing steps performed on the observations analysed in our study. 

Figure \ref{fig10} summarizes information about the error expected for our estimated values by missing the effect of line blending in the analysed data. We notice that this error can lead to either underestimating or overestimating the K$_3$ values depending on the observed region and bandwidth of the analysed data. In particular, for the QS and PL data, it leads to underestimate the K$_3$ parameter up to 2.5~\% for data with 0.7~\AA \ bandwidth, while there is  an overestimation of the values for the data with a spectral degradation in the rage [3,10]~\AA. In this case the error increases linearly up to about 7~\%. On the other hand, for the PO and UM data represented by the corresponding atlas, the approximation applied in our study can lead to an overestimation of K$_3$ values up to 50~\% for data with a bandwidth of 1~\AA, and to an underestimation of the values for data with spectral widths in the range [3,5]~\AA. In this case the underestimation error increases linearly up to about 30~\%. 
Note that the smaller range of bandwidths over which we can investigate the effect of line blending in UM atlas data derives from the fact that those measurements are available only for wavelengths larger than 3920.5~\AA. Overall, the estimated errors suggest that the findings presented above shall be deemed to be indicative of the studied dependence and as an underestimation of its effects. 

Furthermore, we have presented results from spatially-resolved data, both observations and line profiles, as well as from average profiles computed over small areas (20~$\times$~20 pixels wide, i.e. covering a photospheric region of about 0.8$\times$0.8$\arcsec$) selected to represent six solar features in the ambient to which they belong.  We note that the latter data  describe observations that are indeed characterized by a lower spatial resolution than that of the original CHROMIS measurements. This can explain the small dependence on the explored spatial degradations of most of the K$_3$ values reported in Fig. \ref{fig9}. We however acknowledge that those results may not represent findings from similar estimations based on different computational approaches to account for the data characteristics. In particular, we found that the dependence of K$_3$ values on the spectral and spatial degradation of the data slightly differ whether K$_3$ is estimated by maintaining the original resolution of the observations or by degrading it to the one of the selected small areas. The impact of the approach applied to estimate K$_3$ is however minute. In particular,  
Fig. \ref{fig9b} shows the variation of the K$_3$ parameter derived from data of the QS, QM, and QG regions (top panel), and of the PE and UM areas (bottom panel), by considering the original CHROMIS observations and the data obtained from the spectral and spatial kernels applied pixelwise to them. The results in Fig. \ref{fig9b} suggest that the diverse approaches, namely considering the average profile of small areas or profiles derived pixelwise, affect the K$_3$ values obtained, but they only marginally impact the exact form of the relation describing the K$_3$ variation on data characteristics. Indeed, for the ranges of spectral and spatial degradations explored in our study, we found that this relationship is basically  linear over the range [2,10]~\AA \ for all the investigated regions.

Finally, in our study we considered K$_3$ to quantify the \ca brightening in the diverse image pixels and solar features in our data sets, but this quantity is also defined in the literature as line core intensity contrast  
$C$, where $C=I/I_{QS}$, and $I$ and $I_{QS}$ are the Stokes-I intensities at the \ca line core over each image pixel and mean QS intensity averaged over the entire FoV analysed, respectively; see e.g. \citet[][]{Kahil2017}. 

In Fig.~\ref{fig11} we show the variation of the line core intensity contrast estimated on the whole QS region and on the several selected areas of Fig.~\ref{fig2}. Firstly, we notice that the values and trends in Fig.~\ref{fig11} largely differ from the ones reported in Fig.~\ref{fig8} due to the diverse reference intensities employed to derive the quantities in the respective figures. Besides, Fig.~\ref{fig11} displays that the line core contrast decreases monotonically at the increase of the spectral degradation of the data for observations of the whole QS target and UM areas, while it shows a different variation on the selected  areas of QS (QM box 1 and QG box 2) and of PE (box 5) regions. The decrease of the line core contrast in average QS and UM regions is particularly severe for bandwidths up to about 2.5~\AA \ and of the order of 40~\%/\AA \ and 500~\%/\AA, respectively. Results from the selected QS area characterized by regular granulation (QG  box 2) display a hook-like trend with the decrease of the estimated contrast up to a minimum for bandwidths up to about 1~\AA \ and increase of the contrast for larger spectral widths. On the other hand, the contrast values derived from the PE area show a monotonic increase with the bandwidth, which is of the order of 50~\%/\AA \ for observations with bandwidth up to about 5~\AA. 

\section{Summary and Conclusions}
\label{sec5}
We investigated the variation of observations and line profile measurements at the \ca line depending on spectral bandwidth and spatial resolution of the data, and on ambient in the solar atmosphere. We used state-of-the-art observations of the solar photosphere and chromosphere obtained with the CRISP spectropolarimeter and CHROMIS spectrometer operating at the Swedish Solar Telescope. We studied three data sets that relate to very different conditions in the solar atmosphere, representative of a quiet Sun area, a region with plages and several small pores, and a large sunspot with umbra and penumbra. 

Firstly, we noticed the large heterogeneity that characterize the \ca line profiles measured over the studied regions when considered at the full spectral and spatial resolution of the CHROMIS data. This heterogeneity is particularly evident in the data derived from the QS areas, which show intensity variations due to, e.g., $p$-modes oscillations. However, different and  distinguishable elements of the \ca line profiles derived from the diverse studied regions are retained when the data  are averaged over areas selected to represent the solar region to which they belong.     

We then noticed that the effect of spectral degradation on \ca observations and line profiles depends on both the bandwidth employed and observed solar region. As expected, smearing the original data with a spectral kernel leads to greater mixing of photospheric and chromospheric emissions that strongly affects the appearance of the observed regions. In particular, degrading the observations with larger kernels leads to appearance of smaller scale structures and finer details. We found that, among the three regions considered in our study, the one  with plages and several small pores shows the largest change in its appearance with the various spectral degradation kernels applied. 
Moreover, we noticed that the spatial degradation impacts more the data derived from  larger bandwidths  than those obtained with smaller ones. However, the spatial degradation barely affects the appearance of the solar region observed with bandwidths up to 1~\AA, as well as for those observed with a spectral bandwidth in the range [3,10]~\AA. 

Finally, we noticed that the K$_3$ and $EMDX$ parameters employed for the monitoring of the solar and stellar chromospheric activity vary with the spectral bandwidth 
as described by a fifth order polynomial function for all the observed solar regions. 
From the best fitting functions of the data we derived parameters that can be used to intercalibrate results from \ca line observations taken with different instruments in diverse regions of the solar atmosphere.

We would like to emphasize that the line intensities reported from our study  refer to the specific level of solar magnetism and solar activity framed by the analysed observations. Besides, we note that the analysed data are not representative of all the QS, plages, and sunspot regions of the Sun and other stars. Indeed, regions that appear rather similar at the stellar surface may be characterized by a completely different plasma and magnetic topology at higher atmospheric heights, due to e.g. presence of strong field concentrations in the vicinity of the analysed region and affecting it. On the other hand, we note that starting from their common doubly-reversed profile, the \cahk lines share many properties being formed in the chromosphere under similar conditions. We thus expect that the findings from our study on the dependence of the \ca observations and line profiles on characteristics of the data and observed solar ambient   can also qualitatively apply to \cah line data.

\begin{acknowledgments}
The authors thank Vincenzo Andretta for useful comments. 
This study was partly supported by the European Union’s Horizon 2020 research and innovation programme under the grant agreements no. 739500 (PRE-EST project) and no. 824135 (SOLARNET project), the Italian MIUR-PRIN grant 2017APKP7T on “Circumterrestrial Environment: Impact of Sun-Earth Interaction”, and the Italian agreement ASI-INAF 2021-12-HH.0 "Missione Solar-C EUVST -- Supporto scientifico di Fase B/C/D". 

The Swedish 1-m Solar Telescope is operated on the island of La Palma by the Institute for Solar Physics of Stockholm University in the Spanish Observatorio del Roque de los Muchachos of the Instituto de Astrof{\'\i}sica de Canarias. 

This research is supported by the Research Council of Norway, project number 325491 and through its Centers of Excellence scheme, project number 262622.
%
%
%

This study has made use of SAO/NASA Astrophysics Data System's bibliographic services.
\end{acknowledgments}


\appendix

\section{Examples of line profiles}

\begin{figure}
\centering{
\includegraphics[scale=0.14,trim=70 0 0 90, clip]{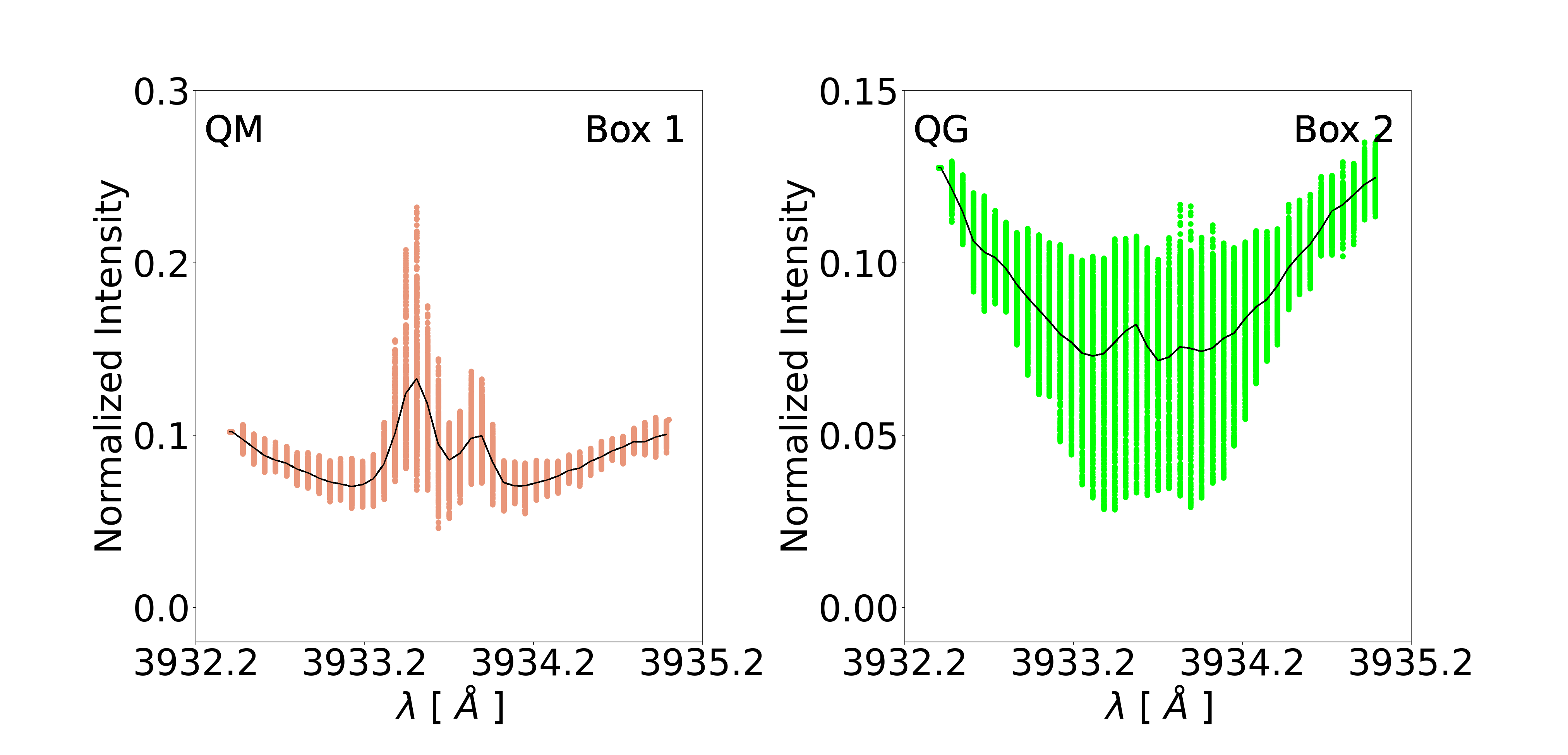}\\
  	\includegraphics[scale=0.14,trim=70 0 0 90,clip]{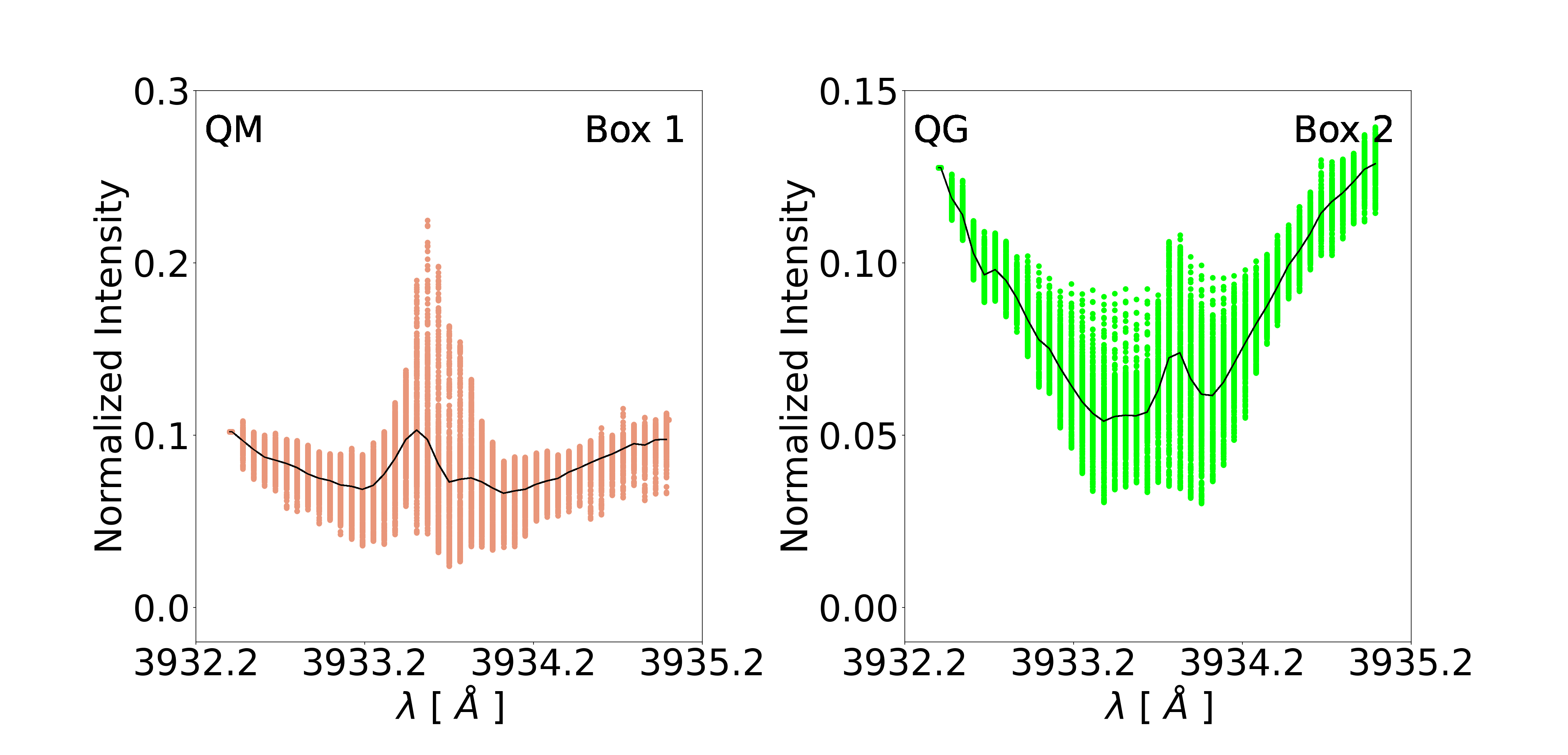}\\
 	\includegraphics[scale=0.14,trim=70 0 0 90, clip]{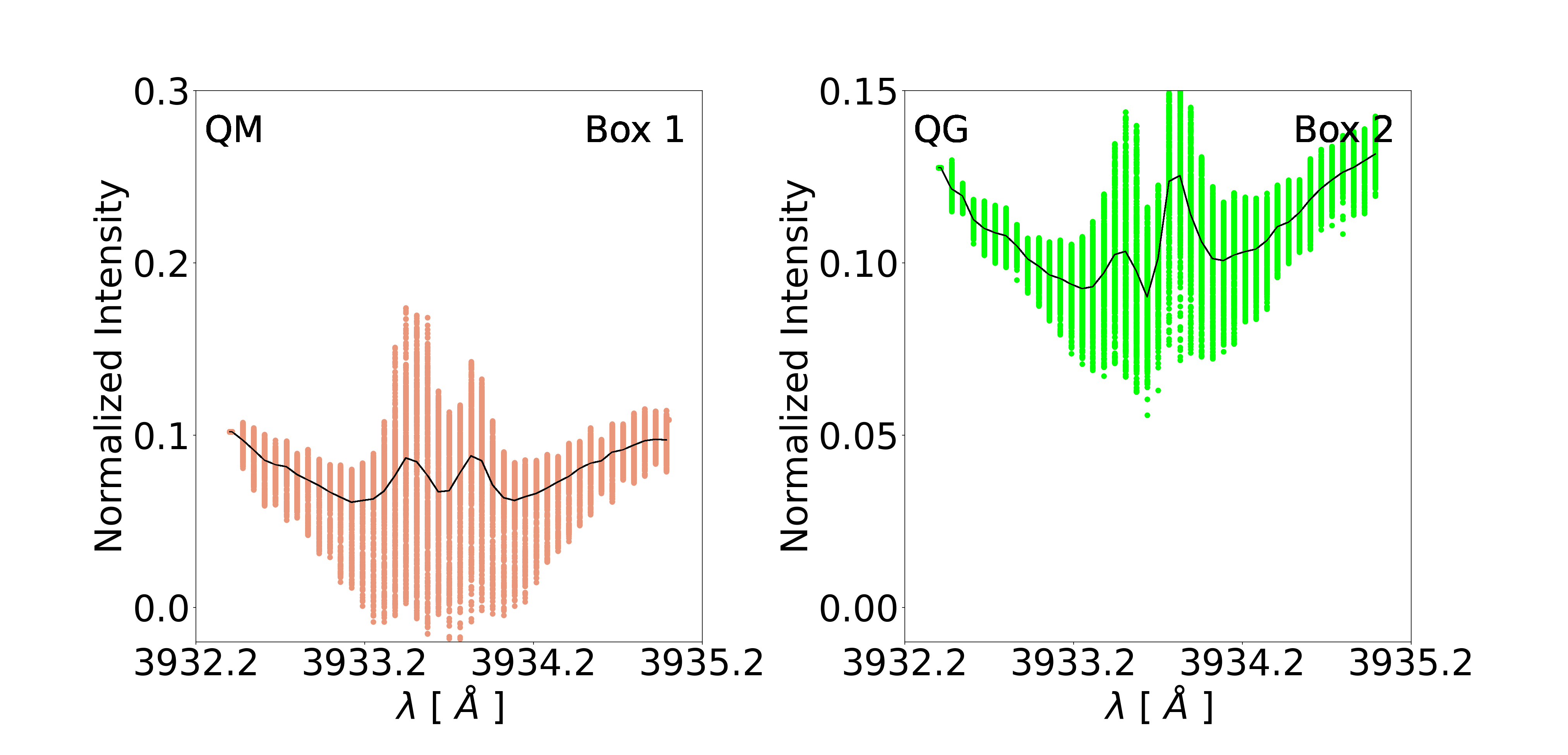}\\
  \includegraphics[scale=0.14,trim=70 0 0 90, clip]{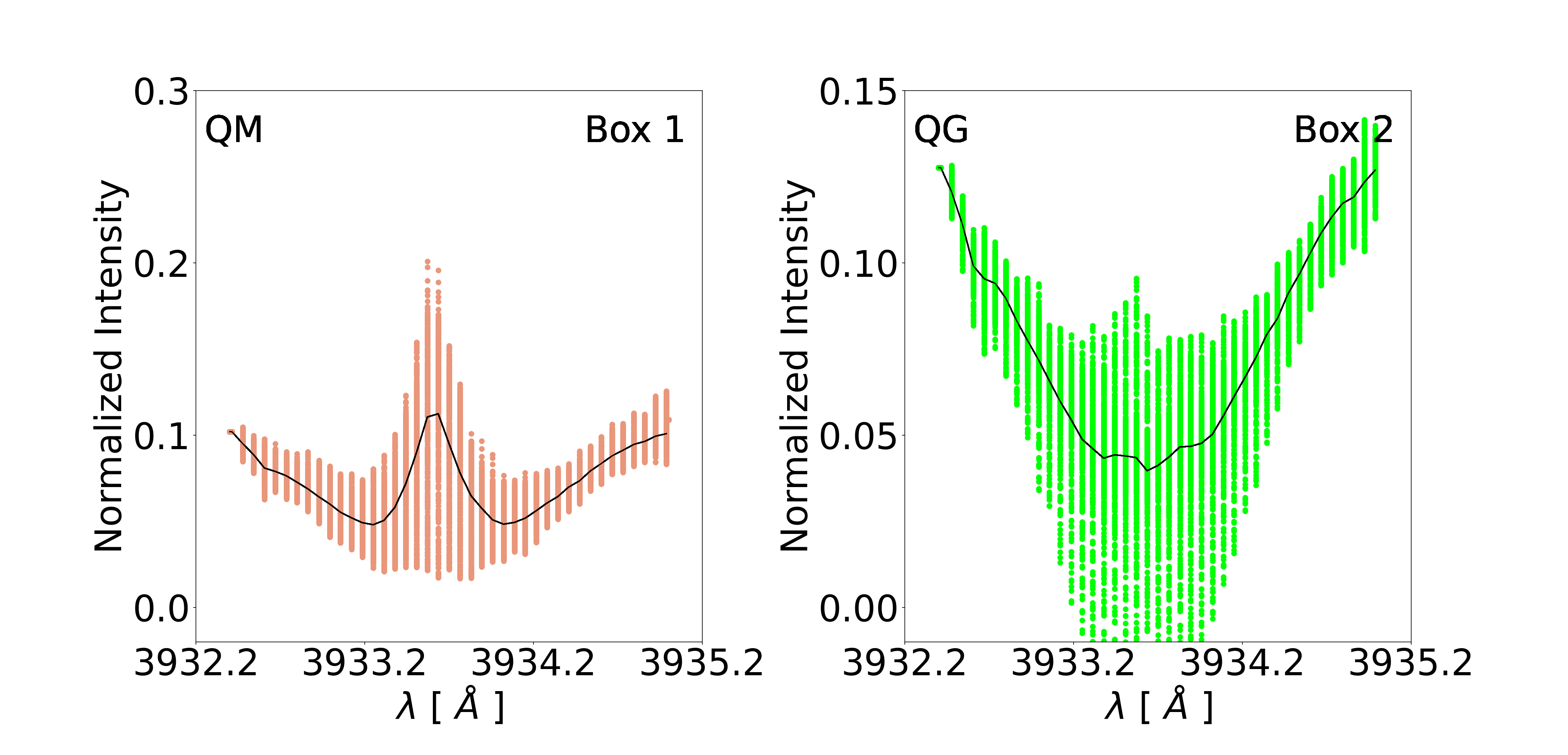}\\
    \includegraphics[scale=0.14,trim=70 0 0 90, clip]{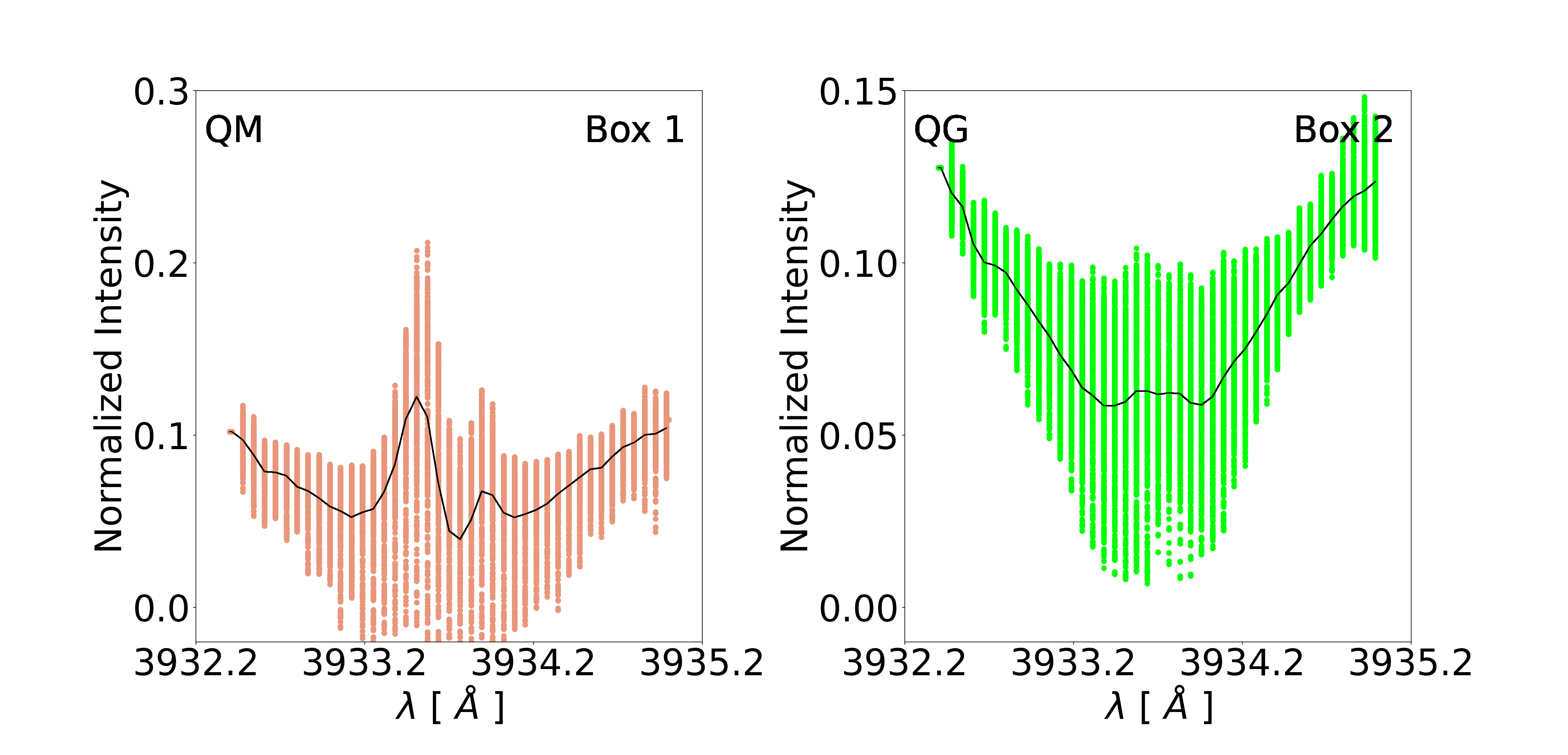}
  }
  \caption{Examples of \ca  line profiles analysed in our study. These were obtained from 20~$\times$~20 pixels wide areas overlapping the  QM and QG boxes marked with numbers 1 and 2 in Fig.~\ref{fig2} or similar adjacent areas. From top to bottom, panels in each row show results derived from boxes marked in Fig.~\ref{fig2}, and from the boxes adjacent to these in their uppermost, lowermost, leftward, and rightward sides, representative of QM (left column panels) and QG (right column panels) regions. Black solid lines show the mean profiles derived from the spatially-resolved data available at each observed spectral position. 
	}
	\label{figa00}
\end{figure}

 Figure \ref{figa00} shows examples of 
line profiles derived from other areas representative of QM and QG features selected nearby the ones reported in Fig.~\ref{fig2}.   
See Sect. 3 for more details.

\section{Examples of spectrally and spatially degraded data}

\begin{figure*}
\centering{
\includegraphics[scale=0.85,trim=0 0 0 0, clip]{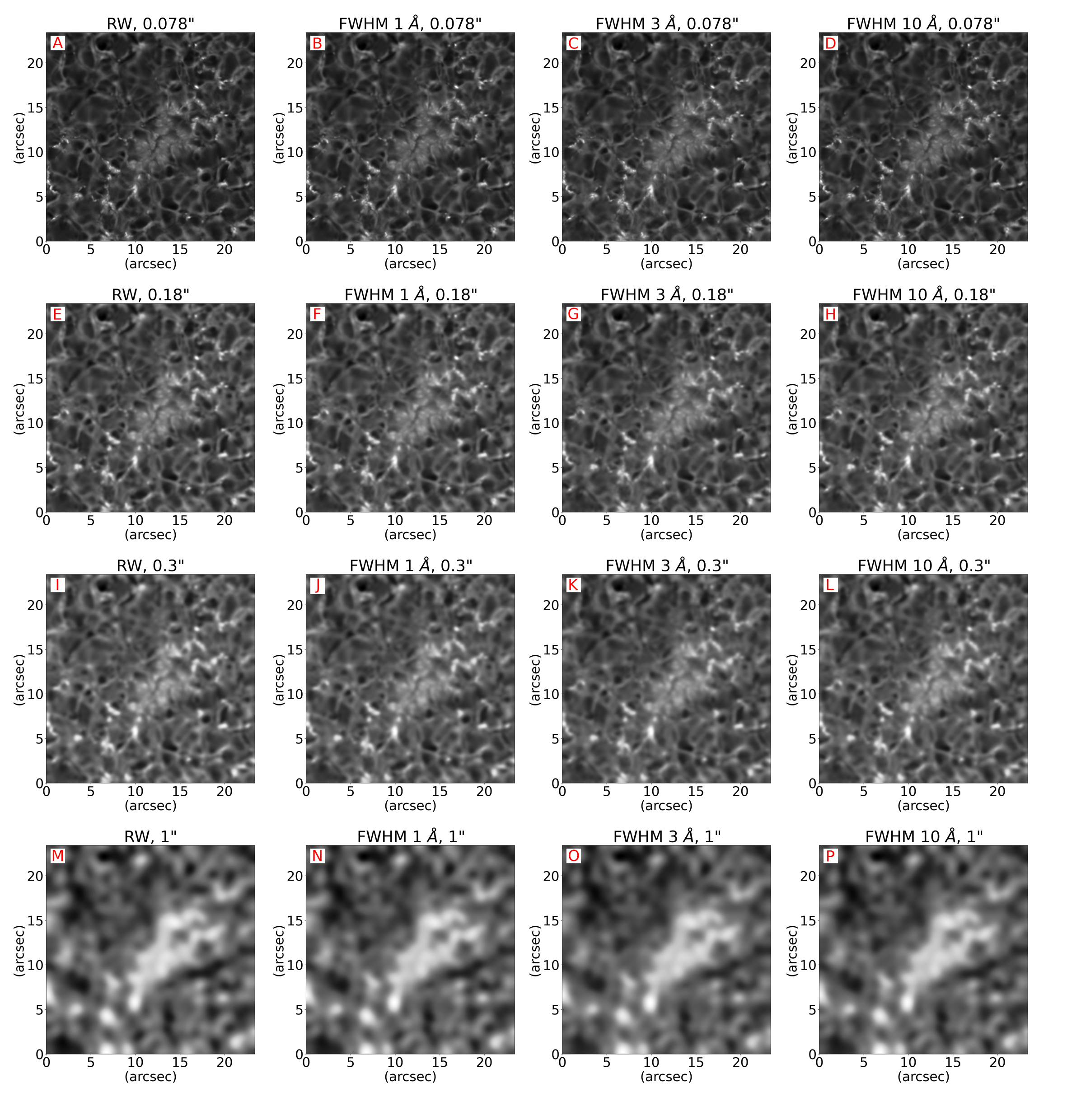}
}
\caption{Examples of the original (panel A) and degraded (all other panels) images of the QS region observed in the red wing (RW) at +1.05~\AA \ from the  \ca  line core, to account for the  diverse bandwidths and spatial resolutions of the most prominent series of available \ca observations. 
Each row shows examples of data characterized by a given pixel scale and by different  bandwidths. From top to bottom, we show data at the original pixel scale of the CHROMIS  observations (panels A--D) and degraded to a spatial resolution  of 
0.18\arcsec \ (panels E--H), 0.3\arcsec \ (panels I--L), and 1.0\arcsec \ (panels M--P), as is in the case of the {\sc Sunrise}/IMaX, Hinode/SOT, and SDO/HMI observations, respectively. For each of these observations, from left to right we show the data at the spectral resolution of the CHROMIS  observations of 0.12~\AA, and the  data spectrally degraded with Gaussian kernels with FWHM of 1, 3, and 10~\AA. Each  image is  shown using the  intensity interval that enhances the visibility of the solar features therein. 
	\label{figa0}
	}
\end{figure*}

\begin{figure*}
\centering{
\includegraphics[scale=0.87]{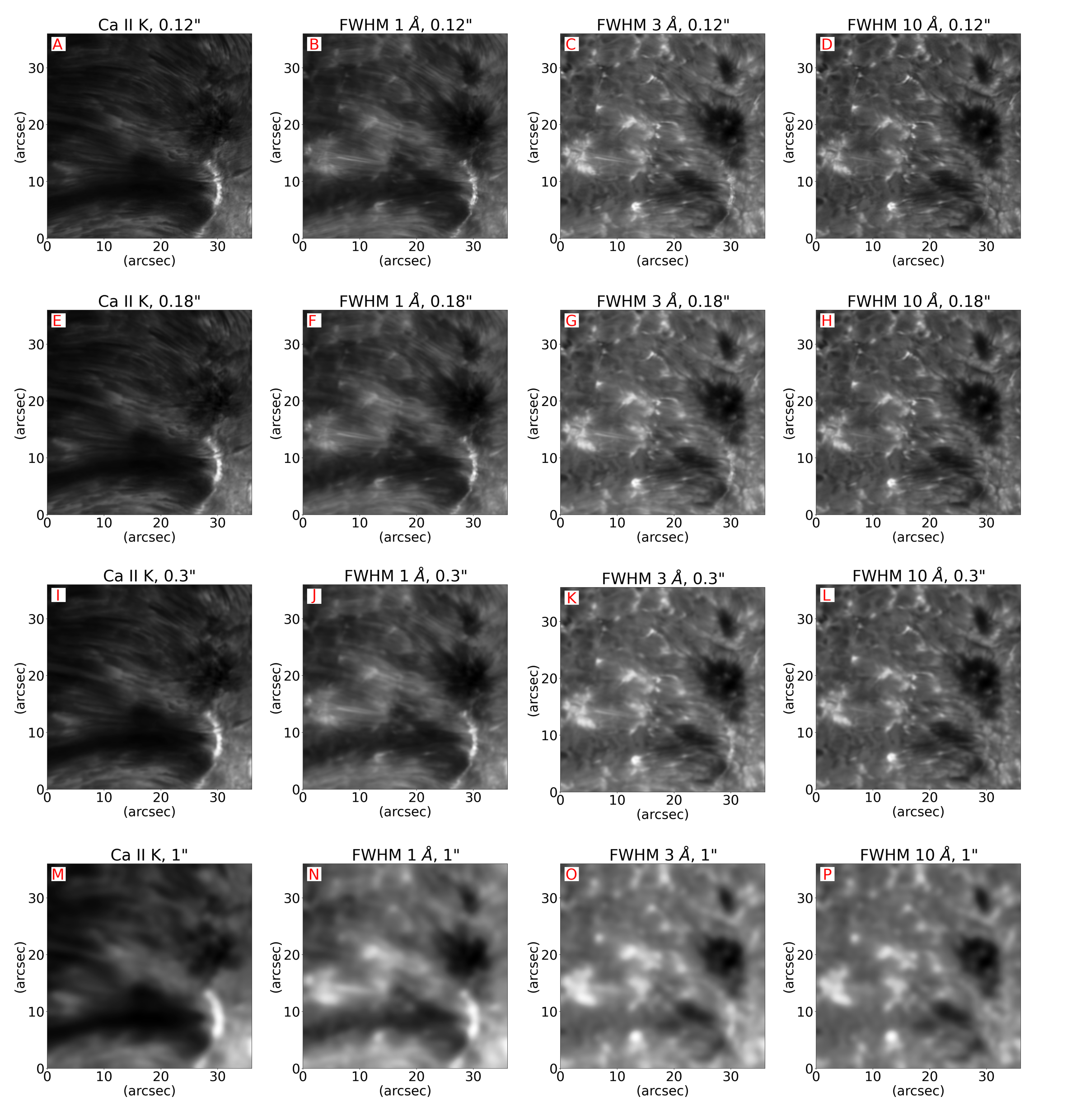}
\caption{Examples of the original (panel A)  and degraded (all other panels) images at the \ca  line core of the NA region including areas with plages (PL) and pores (PO), to account for the  diverse bandwidths and spatial resolutions of the most prominent series of available \ca observations. 
Each row shows examples of data characterized by a given pixel scale and by different  bandwidths. From top to bottom, we show data at the original pixel scale of the CHROMIS  observations (panels A--D) and degraded to a  spatial resolution  of 
0.18\arcsec  \ (panels E--H), 0.3\arcsec  \ (panels I--L), and 1.0\arcsec  \ (panels M--P), as is in the case of the {\sc Sunrise}/IMaX, Hinode/SOT, and SDO/HMI observations, respectively. For each of these observations, from left to right we show the data at the spectral resolution of the CHROMIS  observations of 0.12~\AA, and the  data spectrally degraded with Gaussian kernels with FWHM of 1, 3, and 10~\AA. Each  image is  shown using the  intensity interval that enhances the visibility of the solar features therein.  }
	\label{figa1}
	}
\end{figure*}

\begin{figure*}
\centering{
\includegraphics[scale=0.87]{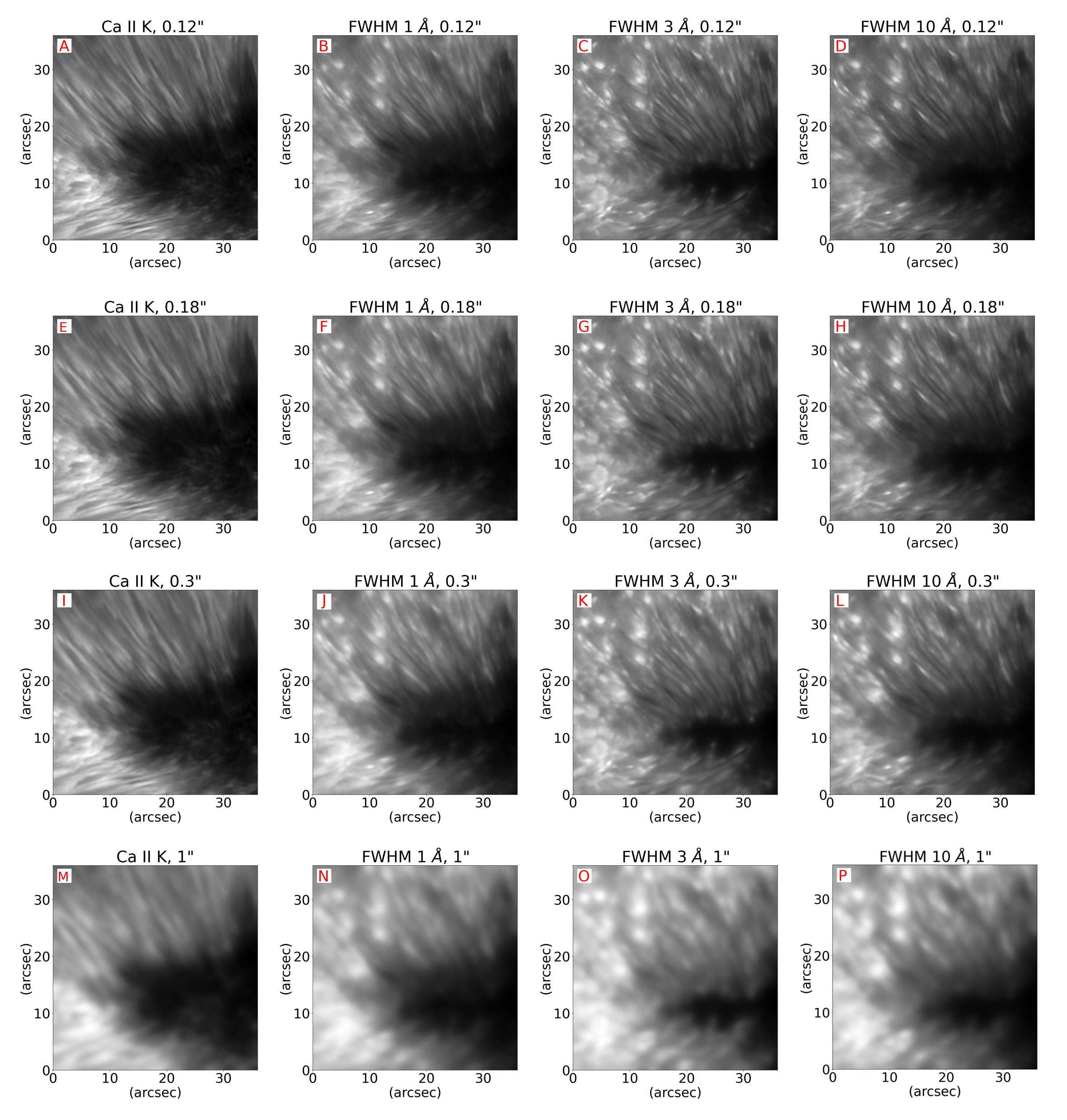}
}
\caption{Examples of the original (panel A)  and degraded (all other panels) images at the \ca  line core of the SP region with UM and PE areas, to account for the  diverse bandwidths and spatial resolutions of the most prominent series of available \ca observations. 
Each row shows examples of data characterized by a given pixel scale and by different  bandwidths. From top to bottom, we show data at the original pixel scale of the CHROMIS  observations (panels A--D) and degraded to a spatial resolution  of 
0.18\arcsec  \ (panels E--H), 0.3\arcsec  \ (panels I--L), and 1.0\arcsec  \ (panels M--P), as is in the case of the pixel scale of the {\sc Sunrise}/IMaX, Hinode/SOT, and SDO/HMI observations,  respectively. For each of these observations, from left to right we show the data at the spectral resolution of the CHROMIS  observations of 0.12~\AA, and the  data spectrally degraded with Gaussian kernels with FWHM of 1, 3, and 10~\AA.  Each  image is  shown using the  intensity interval that enhances the visibility of the solar features therein. 
	\label{figa2}
	}
\end{figure*}

\begin{figure*}
\centering{
\includegraphics[scale=0.87]{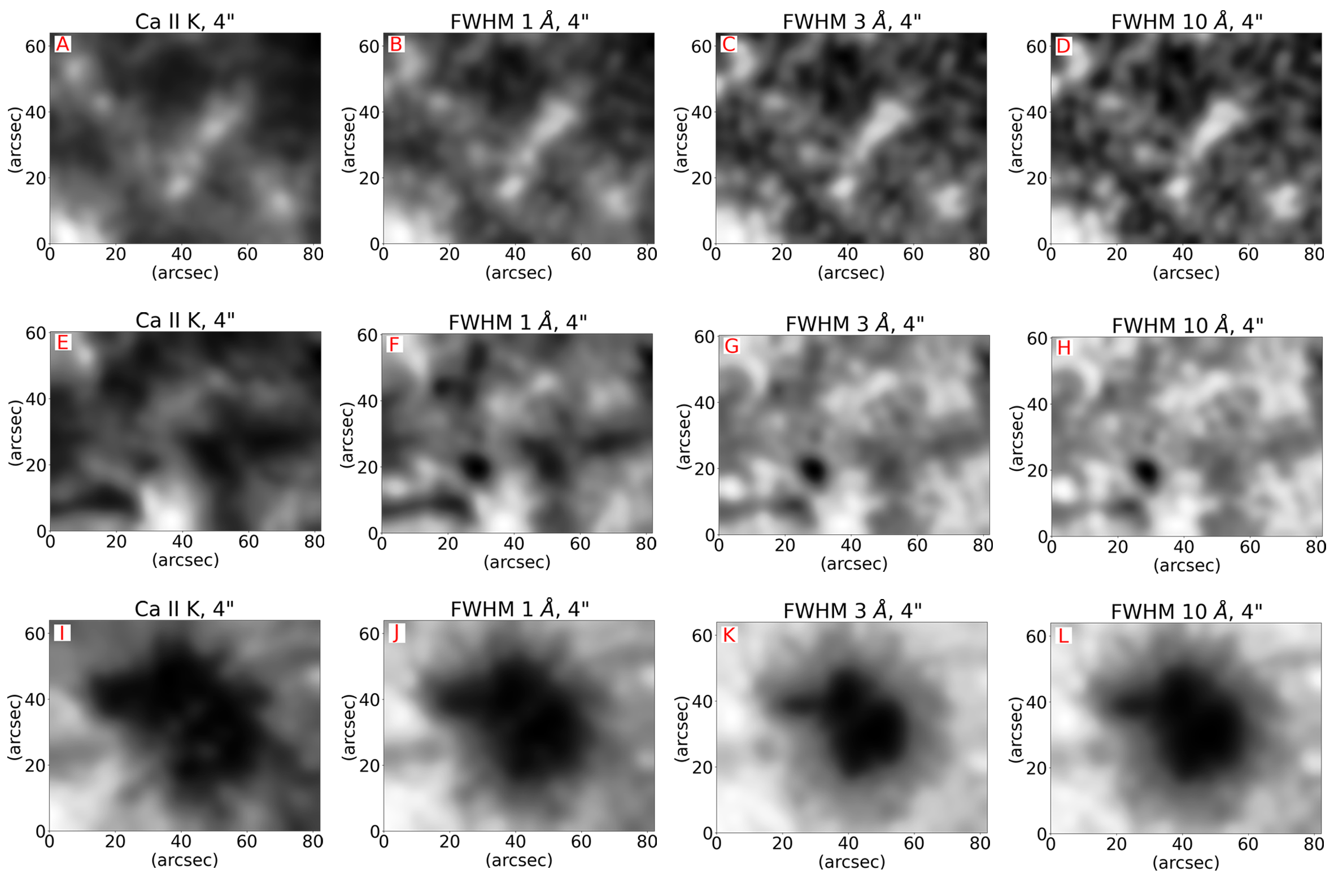}
}
\caption{Examples of 
\ca images in the line core degraded to the moderate spatial resolution of 4\arcsec, which meets the characteristics of several existing series of synoptic full-disc solar observations at the \ca line, of  a quiet Sun area (top row panels), an active region with plages and several pores
(middle row panels), and a sunspot with umbra and penumbra (bottom row panels). From left to right the panels in each row show the three regions as observed at the original spectral resolution of the CHROMIS data (0.12~\AA, panels A, E, I), and with bandwidths of 1~\AA \ (panels B, F, J), 3~\AA \ (panels C, G, K), and 10~\AA \ (panels D, H, L). Each  image is  shown using the  intensity interval that enhances the visibility of the solar features therein. 
} 
	\label{figa3}
\end{figure*}

 Figure \ref{figa0} shows examples of original and  degraded data for the QS region observed  in the red wing of the \ca  line at +1.05~\AA \  from
the line core. Figures \ref{figa1} and \ref{figa2} display examples of original and  degraded data for the NA observations that include plages and pores,  and SP observations with umbral and penumbral regions, respectively. Figure \ref{figa3} shows 
examples of CHROMIS observations in the \ca line core degraded to the moderate spatial resolution of
4~\arcsec \  and spectral degradation given by bandwidths  in the range [0.12,10]~\AA, as 
for the characteristics of most of the existing series of full-disc synoptic solar observations at the \ca line. See Sect. 3.2 for more details.

\end{document}